\documentclass[a4paper, 12pt]{article}  
\usepackage{natbib}
\usepackage{pdfpages}
\usepackage{lmodern}
\usepackage{graphicx} % Required for including pictures 
\usepackage{wrapfig} % Allows in-line images
\usepackage{textcomp} % Allows in-line images
\usepackage[utf8]{inputenc}
\DeclareSymbolFont{letters}{OML}{ztmcm}{m}{it}
\usepackage{color}
\usepackage{bm}
\usepackage{bigdelim, multirow}

\usepackage{array}
\usepackage{listings}
\usepackage{adjustbox}% http://ctan.org/pkg/adjustbox
\usepackage[boxed]{algorithm2e}
\usepackage{multirow} 
\usepackage[OMLmathrm,OMLmathbf]{isomath}
\usepackage{booktabs} 
\usepackage{pdflscape}

\usepackage{comment}
\usepackage[T1]{fontenc} % Required for accented characters
\usepackage{geometry}
\usepackage{tablefootnote}
\usepackage{amsmath}
\usepackage{etoolbox} 
\usepackage{amsthm}
\usepackage{amsfonts}
\usepackage{amssymb}
\usepackage[section]{placeins}
\usepackage{diagbox} 
\usepackage{hyperref}
\usepackage{graphicx} 
\usepackage{fixmath} 
\usepackage[scaled=.90]{helvet}
\usepackage{wrapfig}
\usepackage{tikz}
\usepackage{footnote}
\makesavenoteenv{tabular}
\makesavenoteenv{table}
\usepackage{setspace}
\usepackage[boxed]{algorithm2e}
\usepackage[symbol]{footmisc}
\renewcommand{\thefootnote}{\arabic{footnote}}

\usepackage{multicol}
\usepackage{caption}
    \usepackage[position=top]{subfig}

\usepackage[symbol]{footmisc}

\makeatletter
\renewcommand\@biblabel[1]{\textbf{#1.}} 
\usepackage{enumitem}

\usepackage{pifont}% http://ctan.org/pkg/pifont
\setlist{nolistsep,leftmargin=*}
\renewcommand{\maketitle}{ 
	\begin{center}
		{\LARGE\@title} 
		
		\vspace{0pt} 
		
		{\large\@author} 
		\vspace{30pt} 
	\end{center}
}

\renewcommand{\thefootnote}{\fnsymbol{footnote}} %THIS IS THE FOOTNOTE COMMAND

\title{Estimating excess mortality in high-income countries during the COVID-19 pandemic}
\author{{Giacomo De Nicola$^{1,}\footnote[1]{Corresponding author:  giacomo.denicola@stat.uni-muenchen.de}$, Göran Kauermann$^{1}$}}

\begin{document}

\maketitle
\begin{center}
    
\vspace{-1cm}
{Department of Statistics, LMU Munich, Germany$^1$} \\

\end{center}

\begin{abstract}
%SHORT ABSTRACT:
\noindent{Quantifying the number of deaths caused by the COVID-19 crisis has been an ongoing challenge for scientists, and no golden standard to do so has yet been established. We propose a principled approach to calculate age-adjusted yearly excess mortality, and apply it to obtain estimates and uncertainty bounds for 30 countries with publicly available data. The results uncover considerable variation in pandemic outcomes across different countries. We further compare our findings with existing estimates published in other major scientific outlets, highlighting the importance of proper age adjustment to obtain unbiased figures.}

\vspace{0.2cm}

\end{abstract}

%\newpage %to produce title page

\renewcommand{\thefootnote}{\arabic{footnote}}

\section{Introduction}
%A large number of people have died as a result of the COVID-19 pandemic.

%As´a result of the COVID-19 pandemic, a large number of people have died.

The COVID-19 pandemic has caused a tragically large number of casualties in the general population. Accurately quantifying the magnitude of this number has been a challenge for scientists since the start of the crisis. Knowing how many fatalities were caused by the pandemic is crucial for understanding the factors that govern its spread and severity, and to be able to evaluate the effectiveness of government responses to it. However, death tolls officially related to the virus can only paint an incomplete picture of the situation, as many fatal cases of COVID-19 went undetected in official reports from 2020 and 2021, because of limited testing capacity and misclassification of causes of death \citep{acosta2022global}. Moreover, the reliability of reported deaths varies massively between locations and over time, rendering comparisons largely ineffective. For these reasons, all-cause excess mortality is generally considered to be a more reliable way of assessing the death toll extracted by the pandemic \citep{leon2020covid, beaney2020excess}. 

\begin{figure}[]
    \centering
    \subfloat[\centering Germany]{{\includegraphics[width=0.45\textwidth]{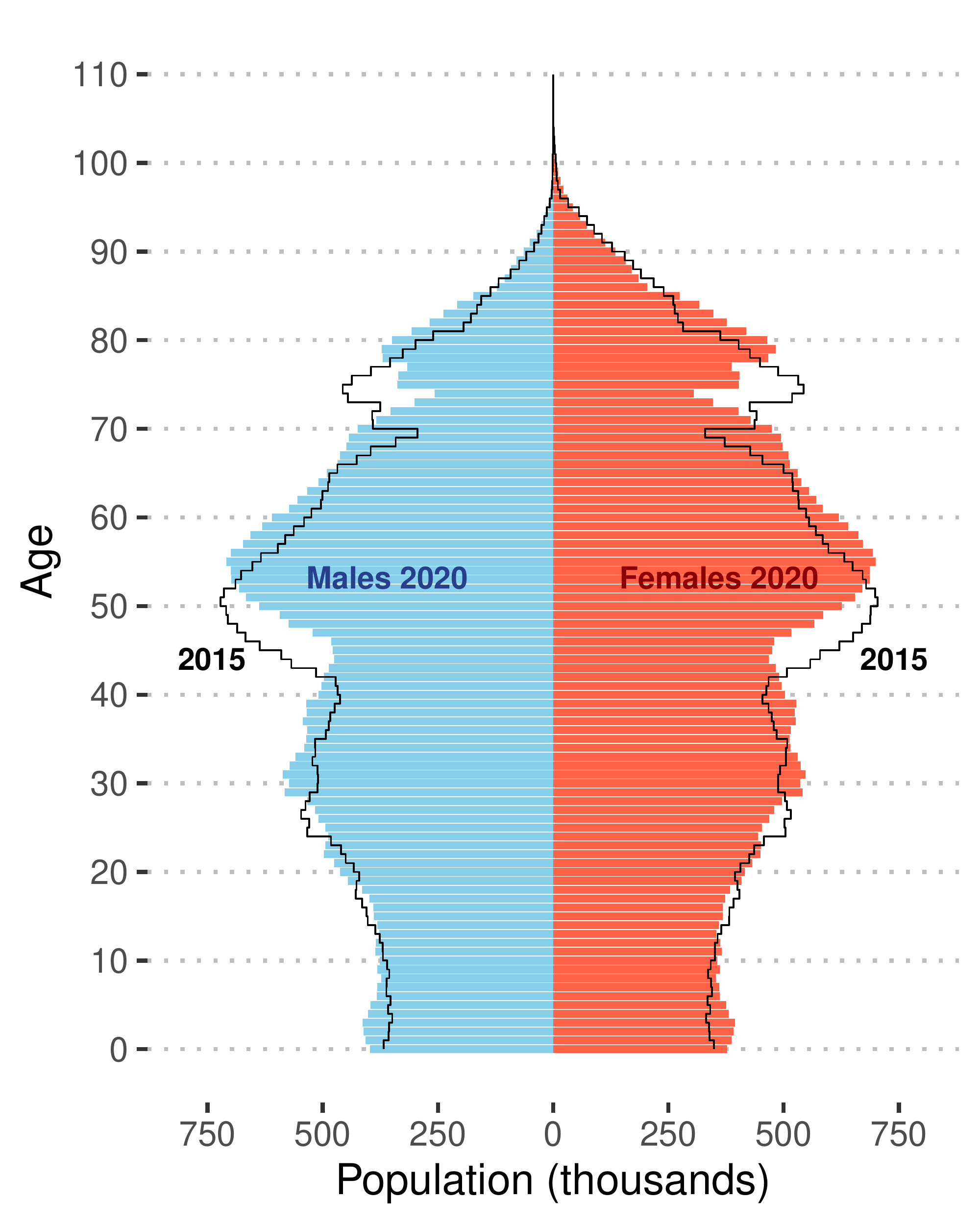} }}%
    %\qquad
    \subfloat[\centering USA]{{\includegraphics[width=0.45\textwidth]{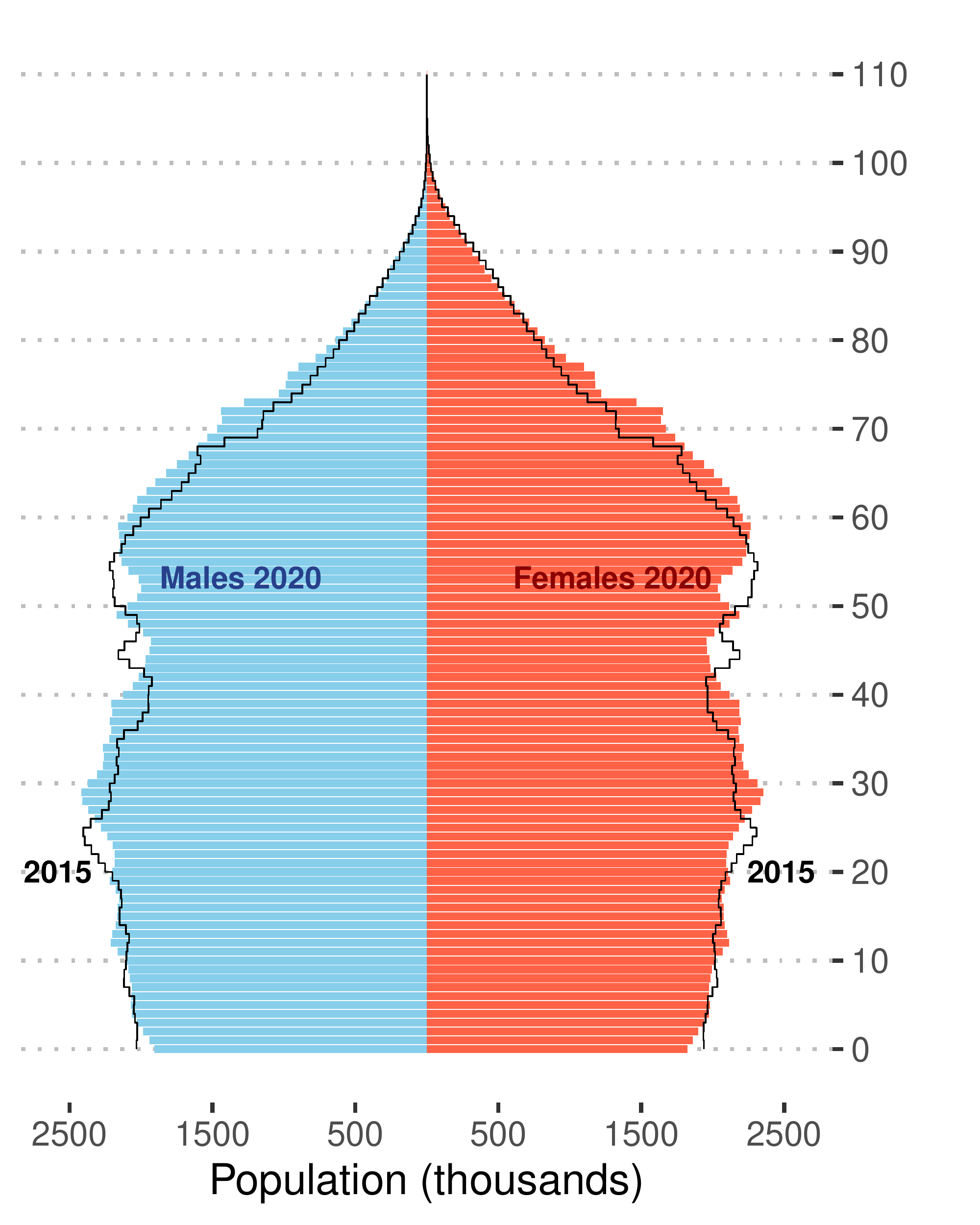} }}%
\caption {Population pyramids in 2015 and 2020 for Germany and the USA. Black and coloured contours indicate population levels in 2015 and 2020, respectively. While the German population is rapidly ageing and its structure is irregular, the USA pyramid is fairly smooth and stable over time.}
\label{fig:pyramids}
\end{figure}

Excess mortality can generally be defined as the number of deaths from all causes during a crisis beyond what we would have expected to see under ``normal'' conditions \citep{checchi2005hpn}. Specifically, our interest here lies in comparing all-cause mortality observed during the COVID-19 pandemic with the overall number of deaths that would have been expected in its absence. If correctly estimated, excess mortality allows to go beyond confirmed COVID-19 deaths, also capturing fatalities that were not correctly diagnosed and reported as well as deaths from other causes that are attributable to the overall crisis conditions.
The concept of excess mortality is well established, and has long being utilised for analysing the impact of wars, natural disasters, and pandemics \citep{johnson2002updating,simonsen2013global}, with its application dating as far back as the Great Plague of London in 1665 (see \citealp{boka2020can}). Today, the concept is routinely employed by governments around the world, with e.g.\ Europe running an early-warning system specifically dedicated to mortality monitoring (the EuroMOMO project, see \citealp{mazick2007monitoring}). Despite this long tradition, however, estimating excess mortality remains a challenge, and no single, unified method for doing so has yet been established \citep{nepomuceno2022sensitivity, acosta2022global}. The difficulty lies in estimating the ``counterfactual'' expected mortality, i.e.\ the number of deaths that would have been expected had the crisis not occurred. This is a hard feat as mortality rates, trends, and data availability vary greatly across different regions and periods of time. Estimating expected mortality thus requires (i) choosing a reference period, and (ii) using some model to project mortality rates from the reference period to the period of interest. This second point also holds for methods that may seem completely ``model-free'', such as the basic approach of simply using the mean number of deaths during the reference period as the expected mortality for the crisis period, widely used in media reports at earlier stages of the pandemic. This method implicitly assumes the total (expected) number of deaths to be constant over both the reference and the crisis period, disregarding other factors that may influence mortality, such as varying life expectancy, due to e.g.\ changes in living conditions, and shifts in the age structure of the population over time. Ignoring the role of age can be particularly damning, as the age structure within a population can change considerably over short periods of time. Moreover, countries can show large variation in how their populations evolve over time, even when their income levels are comparable. To showcase this, Figure \ref{fig:pyramids} depicts the population pyramids of Germany (left panel) and the USA (right panel) during the years 2015 (black contour) and 2020 (coloured), respectively. From the German pyramid, it is apparent how the population aged considerably over the 5-years period prior to the start of the pandemic. In particular, the number of people aged 80 and older increased by approximately 25\% in just five years. Since the probability of death increases sharply at higher ages, the number of deaths in Germany is expected to rise by about 2\% per year as a result of ageing alone, as documented by the German Federal Statistical Office in its annual report \citep{destatis_report}. This means that, for these countries, comparing mortality during the pandemic directly with the reference period will lead to severely underestimating expected mortality, and thus overestimating the excess. %Moreover, given the non-smoothness of the pyramid, and especially due to large ``bump'' and consequent time-variation visible at the ages where risk of death is higher, estimates of expected mortality based on pure trend-fitting also fail at capturing the effect of age. 
If we instead shift our focus to the USA pyramid in the right panel of Figure \ref{fig:pyramids}, we can appreciate how the American population is ageing at a much slower rate than the German one, and that the pyramid is fairly smooth. This means that age adjustment will have a smaller impact on expected and excess mortality estimates. %More importantly, we can also see how the population pyramid for the USA is mostly smooth, and especially so at the older ages: this means that a simple linear trend can do a good job in capturing age-related variation in mortality. 
Unfortunately, the case of the USA is not the norm in modern day high-income countries. To the contrary, what we here showed for Germany is true for many other nations, rapidly ageing as a result of declining fertility rates \citep{bloom2015global}. Due to this, and given that both overall and COVID-related mortality are heavily dependent on age (\citealp{Dowd:2020, Levin:2020}), it is crucial to take age into account to avoid systematic bias in the estimates.

Several high profile studies tackling the estimation of excess mortality in multiple countries try to capture trends in mortality by fitting various regression models to the data in the reference period, and then extrapolating the trend to the period of the crisis to obtain expected mortality figures for that period. \cite{karlinsky2021elife} use linear fits over the years 2015-2019 to obtain excess mortality for 103 countries. The authors use these estimates to compile the World Mortality Dataset \citep{karlinsky2021world}, an important source which is also used by the well know outlet Our World in Data \citep{OWID:2023}. The Economist also runs a page dedicated to tracking excess mortality in the different countries of the world by using a mix of boosted gradient, random forest and bootstrapping \citep{Economist:2023}. \cite{wang2022estimating} provide estimates of excess deaths due to the pandemic in the 2020-2021 period for 191 countries and territories by using an ensemble of six different regression models, including spline regression. These estimates are also the ones officially reported by the IHME \citep{IHME:2022}. While incorporating a trend can account for some of the variation in expected mortality rates over the years, the approach is still not free of problems, as not explicitly accounting for age in the model implicitly assumes %the age structure of the population to evolve at a constant rate over the reference and the study period. In other words, fitting a trend implicitly assumes 
the age pyramid to be smooth. This is often not the case for many modern-day European countries, where demographic traces of World War II are still visible. As an example of this, the German age pyramid depicted in the left panel of Figure \ref{fig:pyramids} clearly shows a bulge in the cohorts entering their ninth decade of life, which causes large non-linear variation in the age-structure of the population over time. For this reason, a trend alone is often not capable of capturing the effect of age.  Furthermore, incorporating country-specific trends in the estimation has the effect of projecting any evolution in the overall death rate observed during the reference period on the period of interest, including those due to factors other than age. While, on the one hand, capturing true long term trends in mortality would be desirable, mortality rates have been shown to exhibit large variance across short periods even in the same region \citep{bergeron2022mortality}. This large variation can lead to predicting large decreases or increases in expected mortality based on variance alone, especially if the trend estimate is based on a period of only 3-5 years, ultimately resulting in overly sensitive and less stable estimates (see \citealp{levitt2023excess} and \citealp{ioannidis2023flaws} for more detail). %On the one hand this is desirable, as it allows to capture true long term trends due to, e.g., improvement or deterioration of living condition. However, mortality rates have been shown to exhibit large variance in different years even in the same region, and they may also differ across different age and gender groups in the same year \citep{bergeron2022mortality}. This large variation can thus lead to predicting large decreases or increases in expected mortality, especially if the trend estimate is based on a period of only 3-5 years, ultimately resulting in overly sensitive and less stable estimates \citep{levitt2023excess}.  %mixes developments in overall mortality due to a changing age structure, and those due to other factors, such as 

Another prominent study is that by \cite{msemburi2022estimates}, which produced the figures officially presented by the WHO \citep{WHO:2022}. The authors use Poisson-based spline regression to fit trends to the reference period, with the exact model specification varying depending on data availability. Excess mortality figures are then obtained by extrapolating a smooth spline based on mortality trends from the five years prior to the pandemic (see \citealp{knutson2022estimating}, for methodological details). While trend extrapolation, as mentioned, can already be problematic when using linear models, the author's extrapolation is based on smoothing splines, which tend to react strongly to short-term fluctuations. Consequently, in countries with weak influenza seasons in 2019, such as many European ones, a steep decline in (expected) mortality is predicted based on a single data point. Extrapolations from smooth splines are generally very sensitive to the last observation (see \citealp{carballo2021general}). Problems with this method have been acknowledged by the authors \citep{van:2022}, who are also working to incorporate age into their analysis and correct their figures \citep{acosta2022global}.

Given the major role that age plays in mortality, many have argued explicit age-adjustment to be a sensible way forward \citep{levitt2022comparison, nepomuceno2022sensitivity, STANG2020797, gianicolo2021gender}.
Several prominent multi-nation studies also do take age into account in their analysis.
\citet{Islamn1137} provide a comparative study on excess mortality in 29 high-income countries in 2020. Their estimates are based on Poisson regression models including the age group (0-14, 15-64, 65-74, 75-85, and >85 years old) as a covariate, and show that different states had excess mortality in different weeks in 2020. %They also point out that countries like New Zealand observed reduced mortality while the highest excess mortality is found in the USA in 2020. 
\cite{konstantinoudis2022regional} also fit Poisson regression models including different age groups, and provide overall figures for five European countries at a fine regional level.
\cite{levitt2022comparison} compare different global excess mortality evaluations, and show the importance of age adjustment by comparing raw calculations with their proposed method for obtaining age-adjusted estimates, showing that the results differ strongly depending on the method used. The author's age adjustment procedure consists in simply dividing the population into the five age strata (the same ones used by \citealp{Islamn1137}) and calculating raw excess mortality for each strata, before summing them up to obtain the overall figure. 

Distinguishing between age classes as done by the studies mentioned above reduces the magnitude of the age-induced bias, and is thus certainly better than not accounting for age at all; on the other hand, simply partitioning the population into age groups is equivalent to assuming age structure to be homogeneous within those age groups, which is unrealistic for, e.g.\, the age group 15-64 in ageing European countries, where the older part of the group (i.e. close to 64 years of age) is decisively more numerous than the younger part (i.e.\ close to 15 years). This is also visible in the German age pyramid depicted in Figure \ref{fig:pyramids}. To eradicate bias from the estimates, it is thus necessary to perform age-standardisation by partitioning the age-pyramid into finer age classes, when data to do so is available. In this paper, we present a simple approach to tackle this issue. Building on  \cite{de2022assessing}, we propose a method to estimate yearly excess mortality figures with fine-grained age adjustment, accounting for any changes occurring in the age structure of the target population over time. Unlike most previously described approaches, our method is not regression-based, and instead directly uses a corrected version of the life tables from the reference period to directly compute expected mortality based on hazards and population by age. This allows to avoid the problems with leverage points mentioned above, and makes the estimates more stable. In addition to point estimates, our approach also allows to compute plausible excess mortality ranges based on the age-specific death rates observed in the years prior to the crisis. 

After showcasing our method, we go on to apply it to obtain excess mortality estimates and ranges for 30 countries for which the necessary data, i.e.\ life tables, population pyramids and yearly deaths tolls, is available. This includes the majority of the world's high income countries. Our results show that, out of these 30 countries, 10 suffered from considerable excess mortality over the years 2020-2021, and 8 displayed a sizeable mortality deficit. In the remaining 12 countries, age-adjusted mortality did not substantially deviate from levels observed during the five years period preceding the pandemic. %We further zoom in on Europe, showing how a spatial effect is visible, with southern and eastern countries being generally more affected than northern ones. 
After presenting our estimates, we compare them with those obtained by five other prominent multi-country studies previously discussed, namely those from the IHME, WMD, Economist, WHO, and \cite{levitt2022comparison}. The comparison showcases the impact of the employed methodology on the results, as the estimates differ greatly based on the method used, and especially so in countries were population is ageing at higher pace. %More specifically, age adjustment matters the most in countries where population is ageing at a higher pace, such as e.g.\ South Korea, Japan and many European countries. To the contrary, results don't substantially differ for countries where the population pyramid is fairly stable, such as, e.g., the USA. 
In addition to presenting our results, we further make data and code to reproduce all of our analyses available in our public GitHub repository \citep{denicola_repository}. This is done to enhance the reproducibility of our results, as well as to facilitate researchers in employing and adapting our methods for further application. 

The remainder of the paper is organised as follows. Section \ref{sec:methods} describes the employed methods, and Section \ref{sec:data} introduces the various data sources used. We present our empirical results in Section \ref{sec:results}, while Section \ref{sec:comparison} is dedicated to the comparison with other prominent studies. %We finally discuss the contributions of this study as well as its limitations in Section \ref{sec:discussion}. 
Section \ref{sec:discussion} concludes the paper with an extended discussion of our contributions.

\section{Methods}
\label{sec:methods}
The main challenge in estimating excess mortality during any crisis lies estimating the ``counterfactual'' expected mortality, i.e.\ the number of deaths that would have been expected had the crisis not occurred. One way to do this is to consider mortality rates observed shortly before the crisis and project them onto the period of interest. A natural approach in this regard is to consider age-specific mortality data contained in official life tables, which give the probability $q_x$ of a person who has completed $x$ years of age to die before completing their next life-year, i.e.\ before their $x+1^{th}$ birthday. The calculation of a life table, as simple as it sounds, is not straightforward, and is an age-old actuarial problem. First references date far back, to \cite{Price:1771} and \cite{Dale:1772}. A historical digest of the topic is provided by \cite{Keiding:1987}. To calculate expected mortality in 2020 and 2021, we here make use of the 2015-2019 5-years period life tables provided by the Human Mortality Database (HMD), one of the most comprehensive and up to date databases on mortality freely available to the public \citep{HMD:2023}. These life tables are calculated as described in Section 7.1 of the HMD Methods Protocol \citep{wilmoth:2021}. The calculation method is akin to that of traditional life tables (see e.g.\ \citealp{Raths:1909}), with the sensible addition of smoothing the death rates via a logistic function for old ages (80+). As demonstrated in \cite{de2022assessing}, further adjustments to the tables are recommendable to relate the expected number of deaths to recently observed ones, especially for countries in which the age pyramid is not smooth, i.e.\ where the assumption of a stationary population (see \citealp{wilmoth:2021}, p.36) does not hold. In particular, population data and life tables need to be appropriately matched, since life tables count the number of deaths of $ x$-year-old people over the course of a year, while population data typically gives the number of $x$-year-old people at a fixed time point (typically the beginning of the year). This requires correction (\ref{eq:qx22}), which accounts for the fact that a person that dies at $x$ years of age in a given year $t$ was either $x$ years old or $x-1$ years old at the beginning of the year, i.e.\ the time point used for the population data. We therefore apply this additional correction, which consists in calculating the adjusted age-specific death probabilities $\tilde{q}_x$ at age $x$ as
\begin{align}
	\label{eq:qx22}
	\tilde{q}_x = \frac{1}{2} q_x +  \frac{1}{2} q_{x+1}\text{,}
\end{align}
where $q_x$ are the death probabilities for age $x$ contained in the life tables before the adjustment. Assuming the maximum possible age to be 110 years, as done in the HMD life tables, we can then compute the overall expected number of deaths in year $t$ as
\begin{align}
	\label{eq:expected2}
	{E}_{t} = \sum_{x=1}^{110} \tilde{q}_x P_{x,t}\text{,}
\end{align}
where $P_{x,t}$ is the population aged $x$ at the beginning of year $t$. We can then obtain excess mortality estimate $\Delta_t$ for a given year $t$ by simply subtracting the expected mortality estimate $E_t$ from the observed death toll $O_t$ in the same year:
$$\Delta_t = O_t - E_t.$$
Computing $\Delta_t$ with $t= 2020, 2021$ using 5-years 2015-2019 life tables yields our excess mortality estimates for the first two years of the COVID-19 pandemic in a given country.

Note that the choice of a 5-years reference period is somewhat arbitrary, but driven by the following principles. In general, it is desirable to have a reference period that is (a) long enough to provide robust data evidence and not fall prey of variance, and (b) short enough to be as similar to the period of interest as possible. Given that the HMD provides life tables calculated on either 1-year, 5-years, or 10-years periods, we picked the 5-years one as it strikes a balance between duration and similarity to the pandemic period. Indeed, using only a single year as the reference period is generally problematic, as yearly death rates exhibit considerable variation, well beyond what can be explained by underlying changes in life expectancy over time. This is particularly evident in our application, as the year 2019, immediately preceding the pandemic, was characterised by relatively low mortality levels in Europe, due to e.g.\ a mild influenza season. On the other hand, using a reference period as long as 10 years is problematic due to the fact that baseline mortality levels can change over such a wide time window. In fact, using data from 2010 to calculate expected mortality 10 years later would downweigh real gains in life expectancy due to e.g.\ changes in living condition and advances in medical technology over time. As such, 5-years life tables seem to be a reasonable choice to strike a balance between bias and variance. We further note that, as shown by \cite{levitt2023excess}, while the choice of the reference period does influence the absolute value of the estimates, it tends not to strongly impact how countries rank relative to one another in terms of excess mortality. In the interest of completeness, we also provide alternative estimates calculated with a 3-years reference period (i.e.\ 2017-2019) and compare them with the 5-years ones in the supplementary material (Section S.3). The comparison highlights how the shorter reference period leads to slightly higher excess mortality estimates, while leaving country rankings largely unaffected. 

An open issue in assessing excess mortality is the quantification of uncertainty. Probability models do not seem very useful here, as variation in mortality is in large part driven by external factors, such as, e.g., the strength of an influenza wave and other exogenous shocks. Because of that, residual variability is well beyond what would be explainable via standard distributional assumptions. %To solve this, one could, in principle, replace the Poisson with a Negative Binomial, or adopt an approach based on quasi-likelihood (see \citealp{mccullagh1983}) and incorporate an additional overdispersion parameter in the model. But in addition to this, stating confidence intervals would also require an understanding of which (super-)population parameters the confidence intervals make statements about, meaning that some kind of repeated sampling setting would have to be assumed \citep{de2022assessing}. 
For this reason, we explicitly refrain from pursuing model-based approaches, and instead propose a data-driven empirical assessment of variability. Specifically, we make use of age-specific single-year mortality rates to provide what we can call a ``plausible range'' for expected mortality. %To do so, we make the assumption that our reference period represents the variation around ``standard'' single-year mortality levels.
To do so, we consider the single-year life tables for each year of the reference period, and use them to calculate expected mortality for the years of interest in the same way as above, i.e.\ using (\ref{eq:qx22}) and (\ref{eq:expected2}). Assuming the reference period to contain a total of $K$ years (in our case $K=5$), we will obtain $K$ different excess mortality estimates. We can then take the lowest and highest resulting estimates as the upper and lower bound of our plausible expected mortality range. In other words, we use mortality rates from the ``worst'' and ``best'' years of the reference period to obtain a plausible range for expected mortality in the years of interest. To be more precise, the upper mortality bound for year $t$ can be written as:
\begin{align}
	\label{eq:expected3}
	{E}_{t}^{\text{upper}} = \max(\tilde{E}_{t,1}, \tilde{E}_{t,2}, ... , \tilde{E}_{t,K}),
\end{align}
where $\tilde{E}_{t,k}$ represents expected mortality for year $t$ calculated using the (corrected) single-year life tables from year $k$. Analogously, the lower bound can be defined as
\begin{align}
	\label{eq:expected4}
	{E}_{t}^{\text{lower}} = \min(\tilde{E}_{t,1}, \tilde{E}_{t,2}, ... , \tilde{E}_{t,K}).
\end{align}
These expected mortality bounds can then be used to obtain excess mortality intervals in a straightforward manner. Note that these bounds do not give a probabilistic measure of uncertainty, as no distributional model is used. Instead, they provide us with plausible high-mortality and low-mortality scenarios for expected mortality in the years of interest based on levels observed during reference years. In a sense, this is akin to the multiverse approach proposed by \cite{levitt2023excess}, whereas instead of presenting all possible universes we only present the average one, the best one and the worst one. %, under the assumption that yearly mortality levels observed in those years can be considered as standard. %In theory, it would indeed be possible to obtain proper confidence intervals by incorporating distributional assumptions in our model. This would, however, not be straightforward for several reasons.

\section{Data}
\label{sec:data}
To compute expected mortality for a given year in a single country/region, our method needs (i) life tables for the reference period and (ii) population data by age for the year of interest. Once expected mortality is calculated, to compute the excess we also need (iii) the yearly death toll for the year of interest. In our case, the period of interest is given by the first two years of the pandemic, i.e.\ 2020 and 2021, while we set the reference period to be 2015-2019. As such, we included in our analysis all countries for which these three pieces of information were readily available at the time of the analysis. We briefly summarise where each of the data pieces was sourced from below, with additional details given in the supplementary material.

\paragraph*{Life tables:}

A great source for population data by year and life tables is given by the Human Mortality Database (HMD), one of the most comprehensive and up to date databases on mortality freely available to the public \citep{HMD:2023}. All life tables used here were sourced from the HMD, and in fact, HMD availability was used as our first inclusion criterion: only countries for which life tables up to 2019 are present in the HMD at the time of the analysis were included in our study. More specifically, 5-year life tables from 2015 to 2019 were used to calculate average expected mortality, while single-year life tables from 2015 to 2019 were used to calculate plausible intervals, as detailed in Section \ref{sec:methods}. Note that all life tables were calculated as described in the HMD's method protocol \citep{wilmoth:2021}, and subsequently adjusted as described in the Methods section.

\paragraph*{Population:}
Population data by single year of age were also sourced from the Human Mortality Database. The presence of this data for the years 2020 and 2021 was the second inclusion criterion for our analysis, as it is needed to calculate expected mortality. Exceptions were made for Italy and Austria, as both countries were central in the COVID-19 debate, especially in the early stages of the pandemic. For both countries, population pyramids were downloaded from the websites of the respective national statistical offices. More details on those sources are given in the supplementary material.

\paragraph*{Deaths:}
Overall death tolls by country for the years 2020 and 2021 are needed to calculate excess mortality in those years. For EU countries, official death tolls were sourced from the Eurostat website \citep{Eurostat:2023}. An exception was made for France, as the Eurostat tolls also include overseas territories; as the HMD life tables refer to mainland France, we sourced mainland death data from the website of the French national statistical office. We also obtained deaths data from the respective official sources for all non-EU countries included in our analysis: details on these sources can be found in the Supplementary Material.

\section{Results}
\label{sec:results}

 Using the method detailed in Section \ref{sec:methods} and following the inclusion criteria detailed in Section \ref{sec:data}, we estimated excess mortality for the years 2020 and 2021 in a total of 30 countries. Table \ref{tab:main} shows all country-specific figures pooled for the years 2020 and 2021. In particular, the table shows, for each country, expected, observed and excess mortality. The table additionally provides the percentage excess mortality, calculated as $\Delta_{t}^{\%}= {\Delta_{t}}/{E_{t}}$, as well as the percentage plausible range for the excess in the two years, calculated as detailed in Section \ref{sec:methods}. Similar tables with separate figures for 2020 and 2021 are given in the Supplementary Material. From Table \ref{tab:main} we can see that the analysed countries exhibit considerable variation in pandemic outcomes, with relative figures ranging from an excess mortality of 22.8\% in Bulgaria, all the way to a mortality deficit of 9.1\% in South Korea. Within our sample, 17 countries had positive excess mortality, while 13 countries saw a mortality deficit during the analysed period. To check whether excess (or deficit) mortality was beyond the expected variation in a given country, we can make use of the estimated plausible range: Indeed, if the range is completely above zero, it means that mortality in the 2020-2021 period was higher than in any of the single years of the reference period, providing solid evidence towards the presence of sizeable excess mortality during the pandemic period. In contrast, a fully negative range implies that mortality during the pandemic period was lower than in any of the years of the 2015-2019 range, indicating a considerable mortality deficit during 2020 and 2021. Using this criterion, we can say that 10 of the 30 analysed countries had substantial excess mortality during the first two years of the pandemic. Those countries, ordered by total absolute excess, are: USA, Italy, UK, Bulgaria, Czechia, Hungary, the Netherlands, Portugal, Croatia, and Lithuania. On the other hand, 8 countries enjoyed a considerable mortality deficit. These are, in order: Japan, South Korea, Taiwan, Australia, Hong Kong, New Zealand, Norway, and Iceland. In all other analysed countries, the two extremes of the range display opposite signs, meaning that pooled mortality from 2020 and 2021 was higher than in the lowest mortality year between 2015 and 2019, but lower than in the highest mortality year in the same range. For these countries, the data thus do not bring conclusive evidence of an excess or a deficit in mortality.

 \begin{table}[h!]
	\centering
	\begin{tabular}{lrrrrrr}
		\hline
		Country & Pop. & Expected & Observed & Excess  &  \%Excess & Plausible range\\ 
		\hline
		Australia & 25.6M & 358,397  &  332,769 & -25,628 & -7.2\% & (-10.4\%, -3.8\%)\\ 
		Austria & 8.9M & 176,736  & 183,561 & 6,825 & +3.9\% & (-0.5\%, +7.1\%)\\ 
		Belgium & 11.5M & 232,757 & 239,227 & 6,470  & +2.8\% & (-2.1\%, +7.4\%)\\ 
		Bulgaria & 6.9M & 222,890 & 273,730 & 50,830 & +22.8\% & (+19.5\%, +25.5\%)\\ 
		Canada & 37.9M & 611,142 & 620,052 & 8,910 & +1.5\% & (-0.3\%, +4.1\%)\\ 
		Croatia & 4.0M & 111,092 & 119,735 & 8,642 & +7.8\% & (+1.9\%, +12.9\%)\\ 
		Czechia & 10.5M & 237,489 & 269,180 & 31,691 & +13.3\% & (+8.7\%, +16.7\%)\\ 
		Denmark & 5.8M & 115,349 & 111,797 & -3,552 & -3.1\% & (-5.2\%, +0.1\%)\\ 
  	Finland & 5.5M & 116,484  & 113,147 & -3,337 & -2.9\% & (-5.4\%, +1.0\%)\\ 
		France & 65.4M & 1,268,041  &  1,298,800 & 30,759 & +2.4\% & (-0.5\%, +5.3\%)\\ 
		Germany & 83.2M & 2,005,161 & 2,009,259 & 4,098  & +0.2\% & (-3.2\%, +3.5\%)\\ 
		Hong Kong & 7.5M & 108,628 & 102,189 & -6,439 & -5.9\% & (-12.1\%, -1.4\%)\\ 
		Hungary & 9.8M & 270,804 & 297,457 & 26,653 & +9.8\% & (+6.0\%, +13.0\%)\\ 
		Iceland & 0.4M & 4,926 & 4,637 & -289 & -5.8\% & (-11.2\%, -2.0\%)\\ 
		Ireland & 5.0M & 68,404 & 65,445 & -2,959 & -4.3\% & (-8.6\%, +0.4\%)\\ 
		Italy & 59.4M & 1,352,461 & 1,449,352 & 96,891 & +7.2\% & (+2.0\%, +10.8\%)\\ 
  	Japan & 123.5M & 2,944,310  &  2,825,044 & -119,266 & -4.1\% & (-6.4\%, -1.9\%)\\ 
		Lithuania & 2.8M & 83,293  & 91,293 & 8,000 & +9.6\% & (+3.5\%, +17.2\%)\\ 
		Luxembourg & 0.6M & 9,211 & 9,098 & -113  & -1.2\% & (-3.0\%, +2.1\%)\\ 
  	Netherlands & 17.4M & 325,475 & 339,650 & 14,175  & +4.4\% & (+1.9\%, +7.9\%)\\ 
		New Zealand & 5.1M & 72,746 & 67,545 & -5,201 & -7.2\% & (-9.3\%, -5.4\%)\\ 
		Norway & 5.4M & 87,031 & 82,613 & -4,418 & -5.1\% & (-8.2\%, -1.6\%)\\ 
		Portugal & 10.3M & 237,744 & 248,198 & 10,454 & +4.4\% & (+1.9\%, +7.8\%)\\ 
		South Korea & 51.3M & 684,662 & 622,628 & -62,034 & -9.1\% & (-14.4\%, -2.6\%)\\ 
		Spain & 47.4M & 905,407 & 941,717 & 36,310 & +4.0\% & (-1.2\%, +9.4\%)\\ 
        Sweden & 10.4M & 192,763 & 190,082 & -2,681 & -1.4\% & (-4.2\%, +4.1\%)\\ 
        Switzerland & 8.6M & 145,131 & 147,387 & 2,256 & +1.6\% & (-4.3\%, +5.2\%)\\ 
        Taiwan & 23.6M & 383,471 & 357,239 & -26,232 & -6.8\% & (-10.4\%, -3.2\%)\\ 
        UK & 66.9M & 1,285,300 & 1,357,108 & 71,808 & +5.6\% & (+2.4\%, +9.5\%)\\ 
        USA & 330.7M & 5,921,695 & 6,842,426 & 920,731 & +15.6\% & (+14.1\%, +17.4\%)\\ 
		\hline
	\end{tabular}
	\caption{Expected and observed yearly mortality in the 2020-2021 period for each of the 30 countries included in the analysis.}
	
	\label{tab:main}
\end{table}

\begin{figure}[]
	\centering
 \includegraphics[width = 0.9\textwidth]{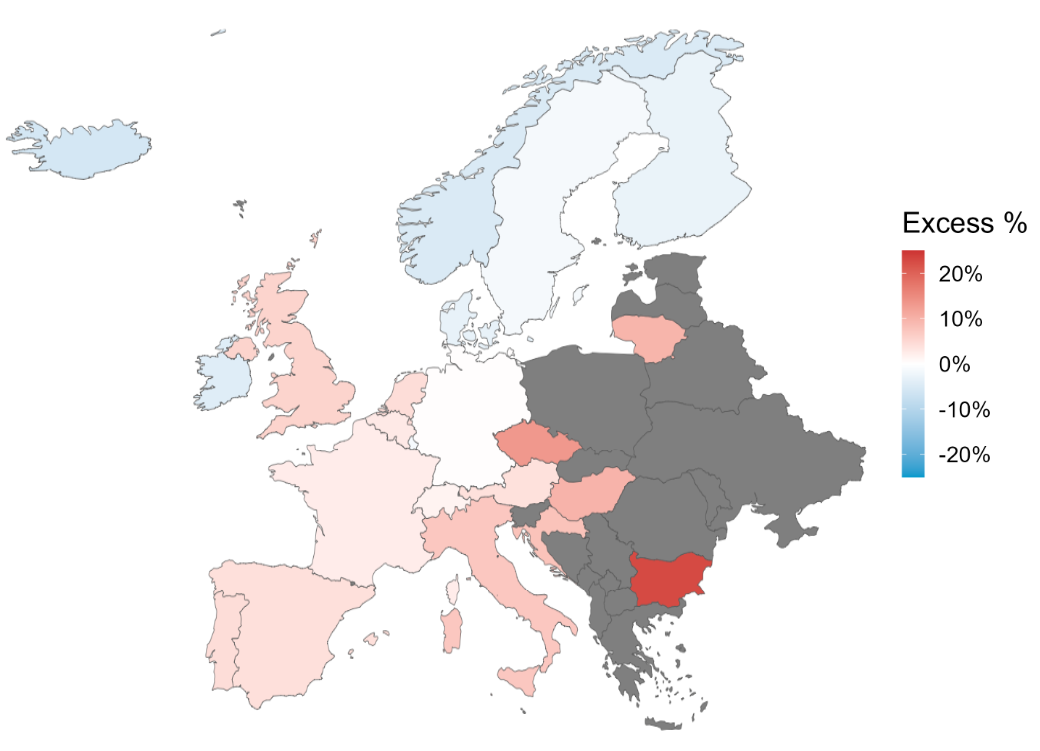}

\caption {Heat-map of excess mortality in Europe in the 2020-2021 period. Dark grey indicates that the necessary data is not available at the time of the analysis.}
\label{fig:map}
\end{figure}

As mortality data is inherently spatial in nature, it is often useful to visualise it on a map, to allow  for a better overall view as well as to recognise any spatial patterns that may emerge. Given that many of the analysed countries are in Europe, and given that our data cover most of the western part of the continent, we provide a heat-map of excess mortality in Europe during the 2020-2021 period in Figure \ref{fig:map}. %The colour coding uses red for indicating excess mortality and blue for mortality deficit, with darker shades representing more extreme values.
From the map, we can clearly see that Bulgaria stands out as the state with the highest excess mortality, with a value of 22.8\% (as seen from Table \ref{tab:main}). 
%maybe following for discussion?
%----
An important thing to note in this regard is that Bulgaria is the only country within our sample which is not high-income by World Bank standards, but is rather found within the upper-middle income bracket \citep{worldbank:2022}. This is relevant, as it provides an indication of how much harder lower income countries were hit by the pandemic in terms of life loss. While our study only focuses on countries for which the necessary data is fully available, i.e.\ primarily high-income countries, studies working with incomplete data such as those of \cite{msemburi2022estimates} and \cite{karlinsky2021elife} corroborate this.
%----
From the map, a spatial pattern is also visible: Within Europe, southern and eastern countries suffered from excess mortality, while northern countries mostly display mortality deficits. Lower mortality in the Nordic countries may be due to a combination of campaigns delivering vaccines faster to more people than the European Union (EU) average, effective non-pharmaceutical public health interventions (NPIs) and high baseline capacities of the health care systems \citep{scholey2022life}. Lower population densities may also have played a role in stifling the spread of the disease \citep{rocklov2020}.

\begin{figure}[]
	\centering
 \includegraphics[width = 0.496 \textwidth]{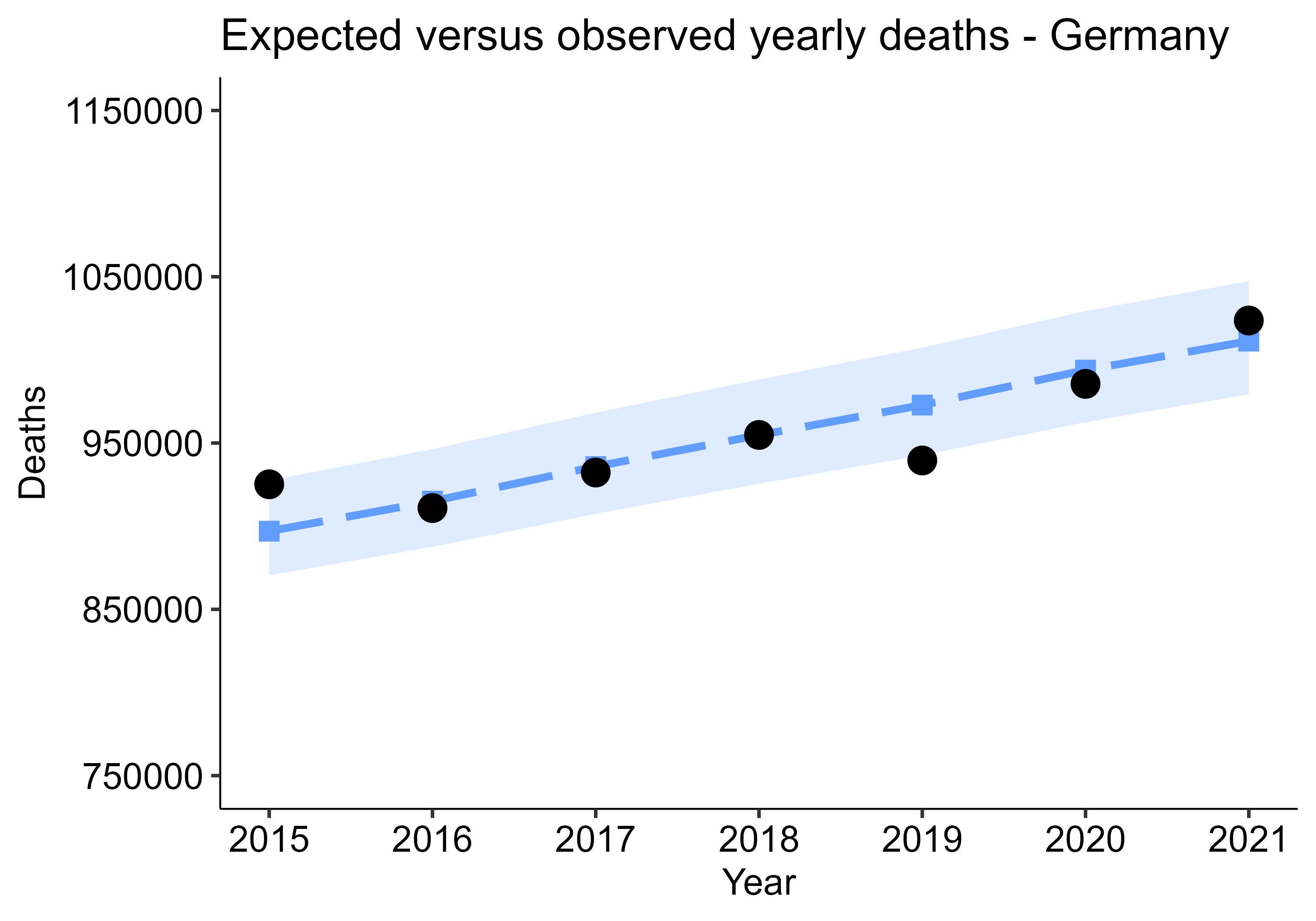} 
\includegraphics[width = 0.496 \textwidth]{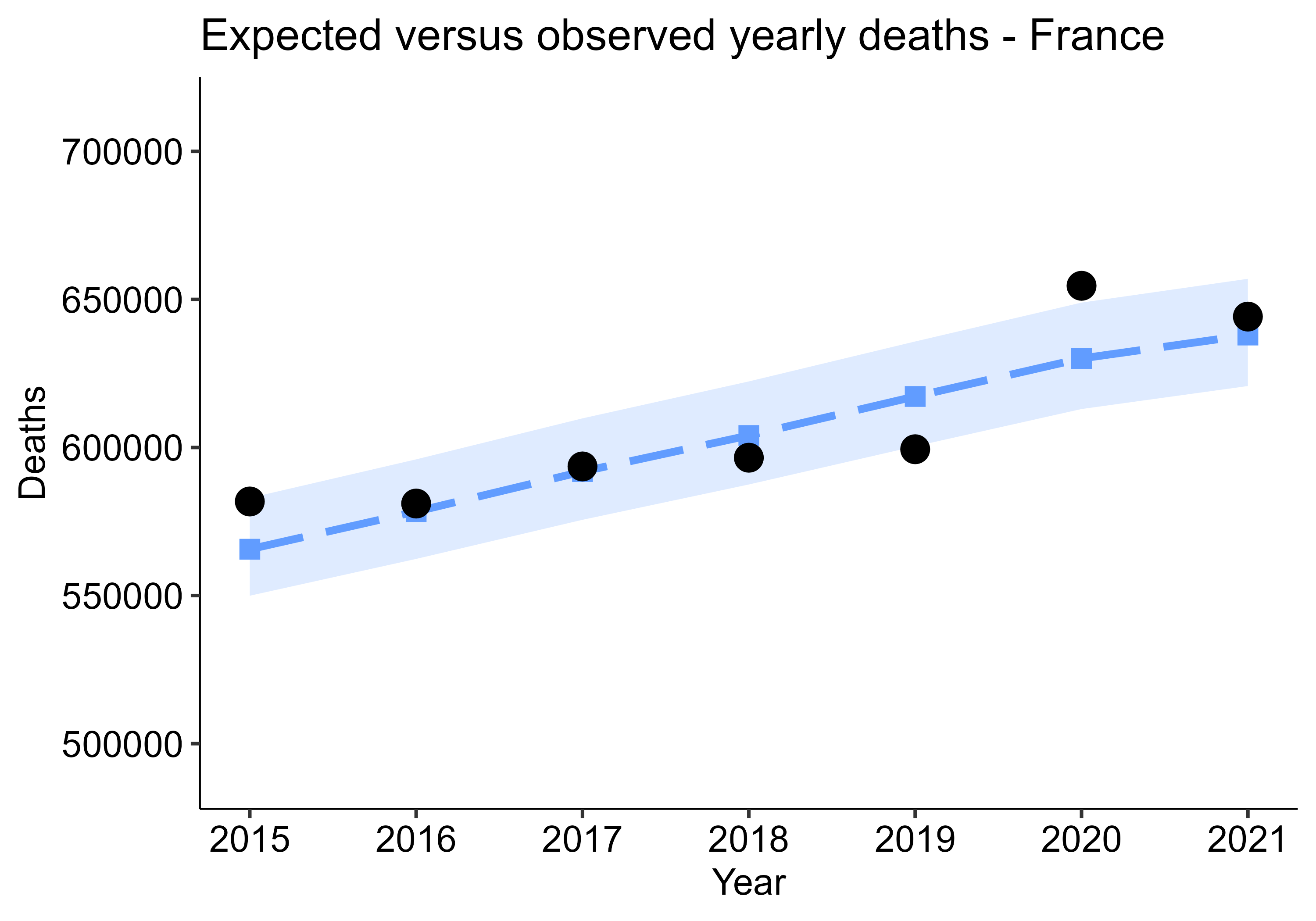}
\includegraphics[width = 0.496 \textwidth]{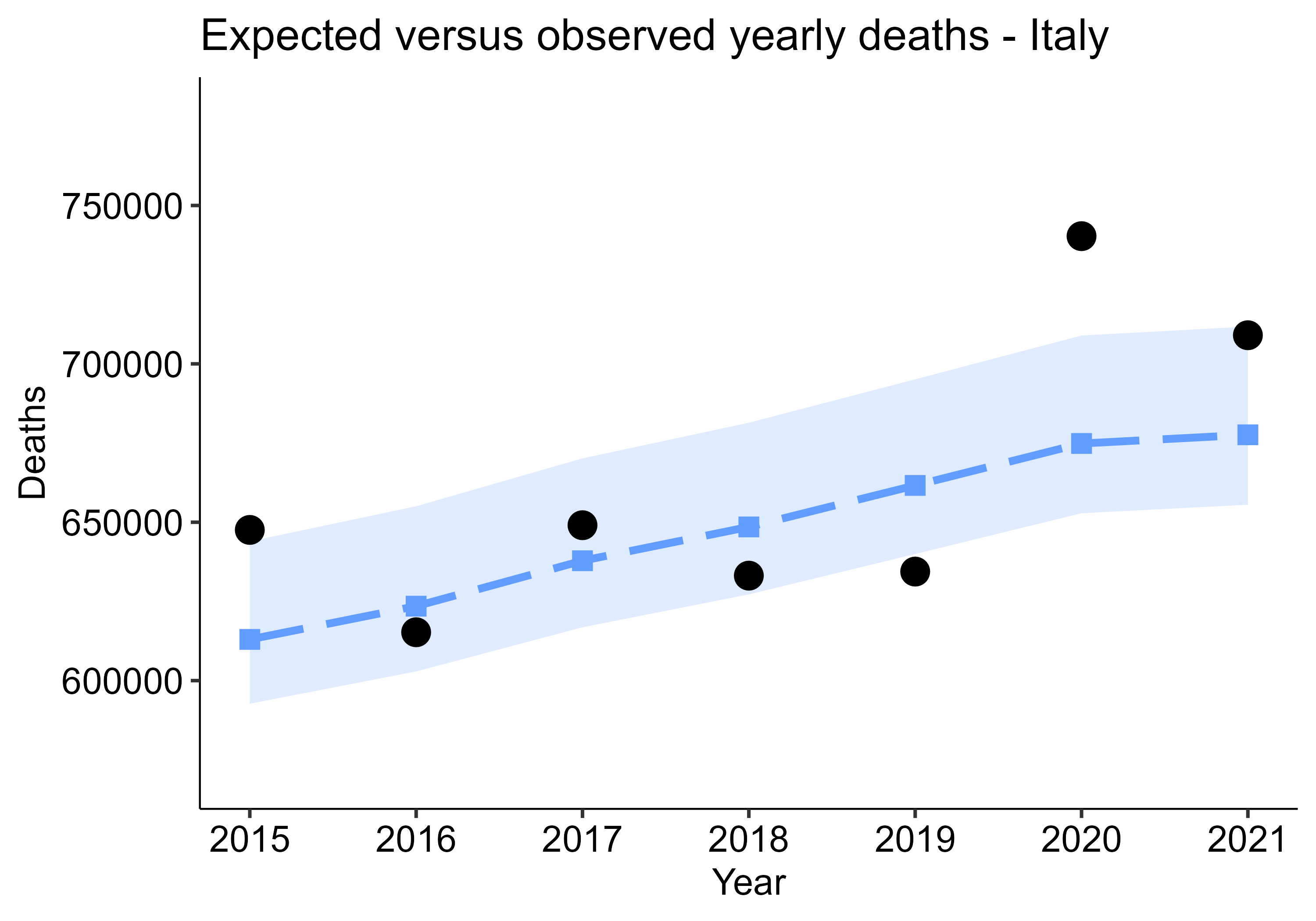}
\includegraphics[width = 0.496 \textwidth]{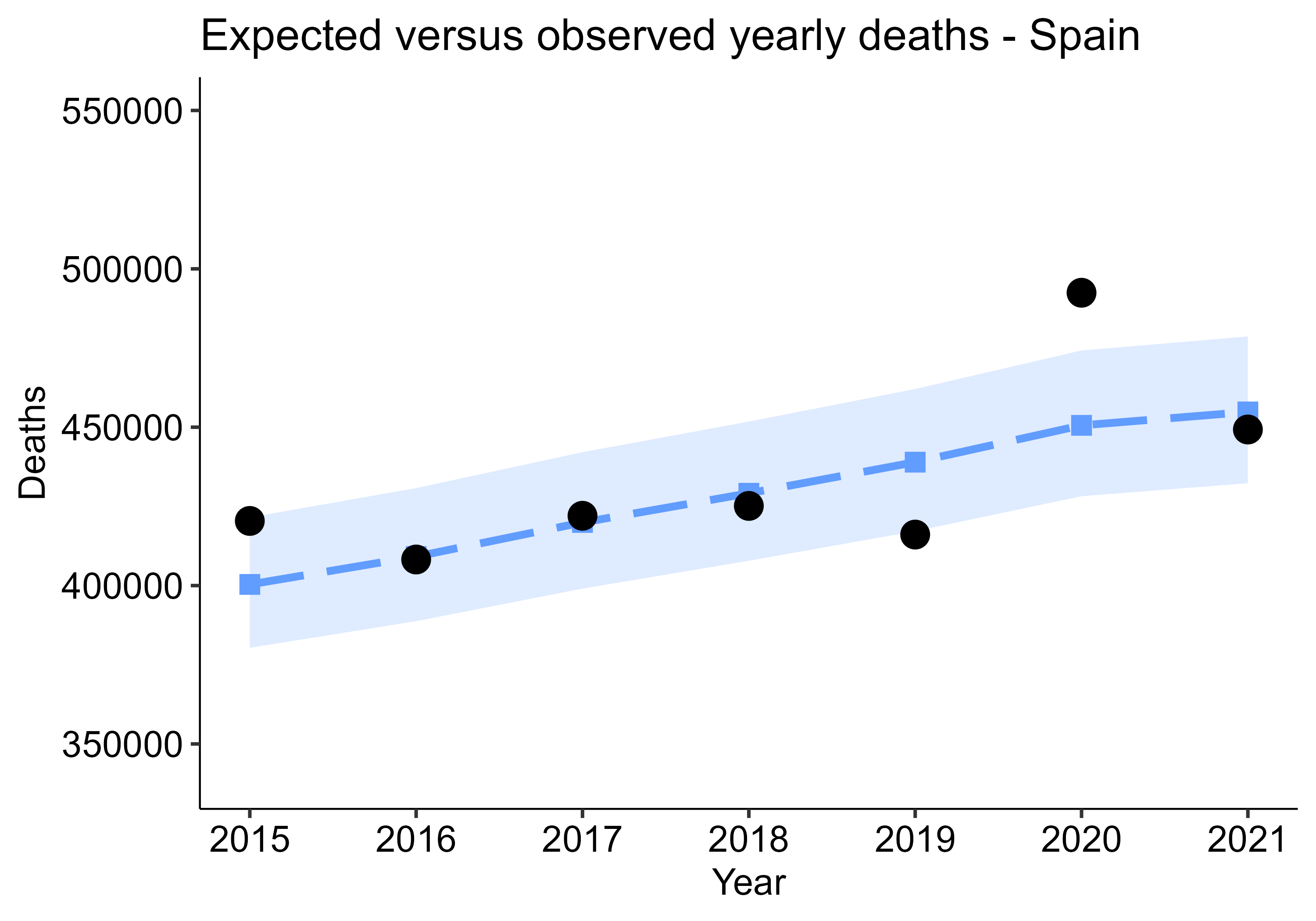}

\caption {Expected and observed mortality figures by calendar year for the four largest EU countries: Germany, France, Italy and Spain. Black dots indicate observed mortality in a given calendar year, while blue squares indicate estimated yearly expected mortality. Shaded bands represent the estimated expected mortality range.
}
\label{fig:EU}
\end{figure}

Zooming in on the four largest EU countries, Figure \ref{fig:EU} plots expected and observed mortality figures by calendar year in Germany (top-left), France (top-right), Italy (bottom-left) and Spain (bottom-right). Note that expected deaths from years 2015-2019 were calculated the same way as the 2020-2021 ones, i.e.\ using corrected 2015-2019 life tables. We plot the observed death counts for each year as black dots, while expected death counts are represented by blue squares. Furthermore, the light-blue bands mark the plausible expected mortality range for each year, calculated using (\ref{eq:expected3}) and (\ref{eq:expected4}). From the figure we can immediately appreciate the importance of age adjustment for these four countries, as both expected and observed mortality tend to naturally increase year over year due to ageing populations. Calculating excess mortality using raw data would thus return inflated figures. Focusing on the single countries we can see how Germany did not suffer from sizeable excess mortality during the pandemic, with age-adjusted mortality in 2020 and 2021 being in line with previous years. On the other hand, the three Mediterranean countries depicted all suffered from varying degrees of increased mortality in 2020, with mortality converging back to more normal levels in the second pandemic year.

\begin{figure}[h!]
	\centering
 \includegraphics[width = 0.496 \textwidth]{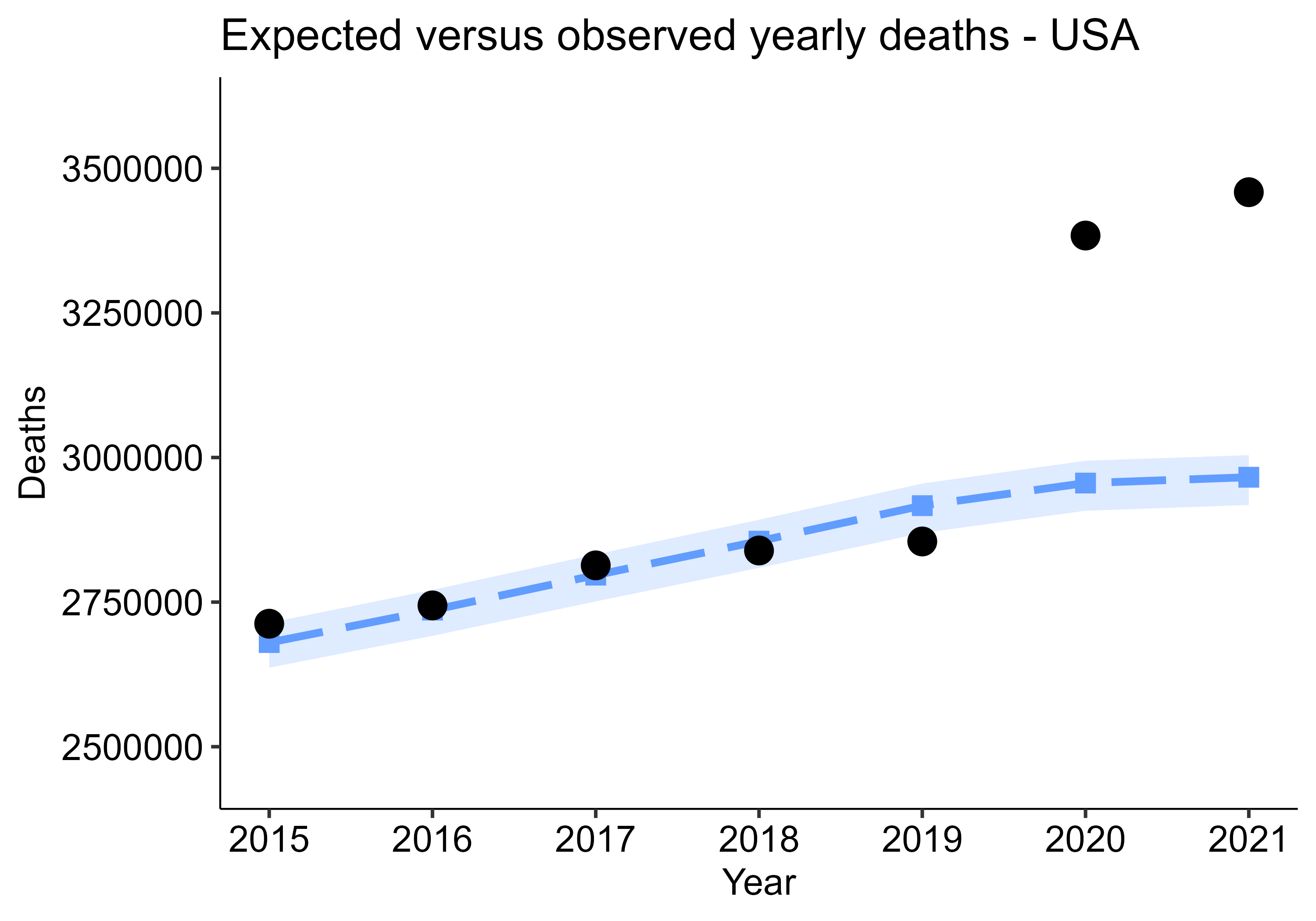} 
\includegraphics[width = 0.496 \textwidth]{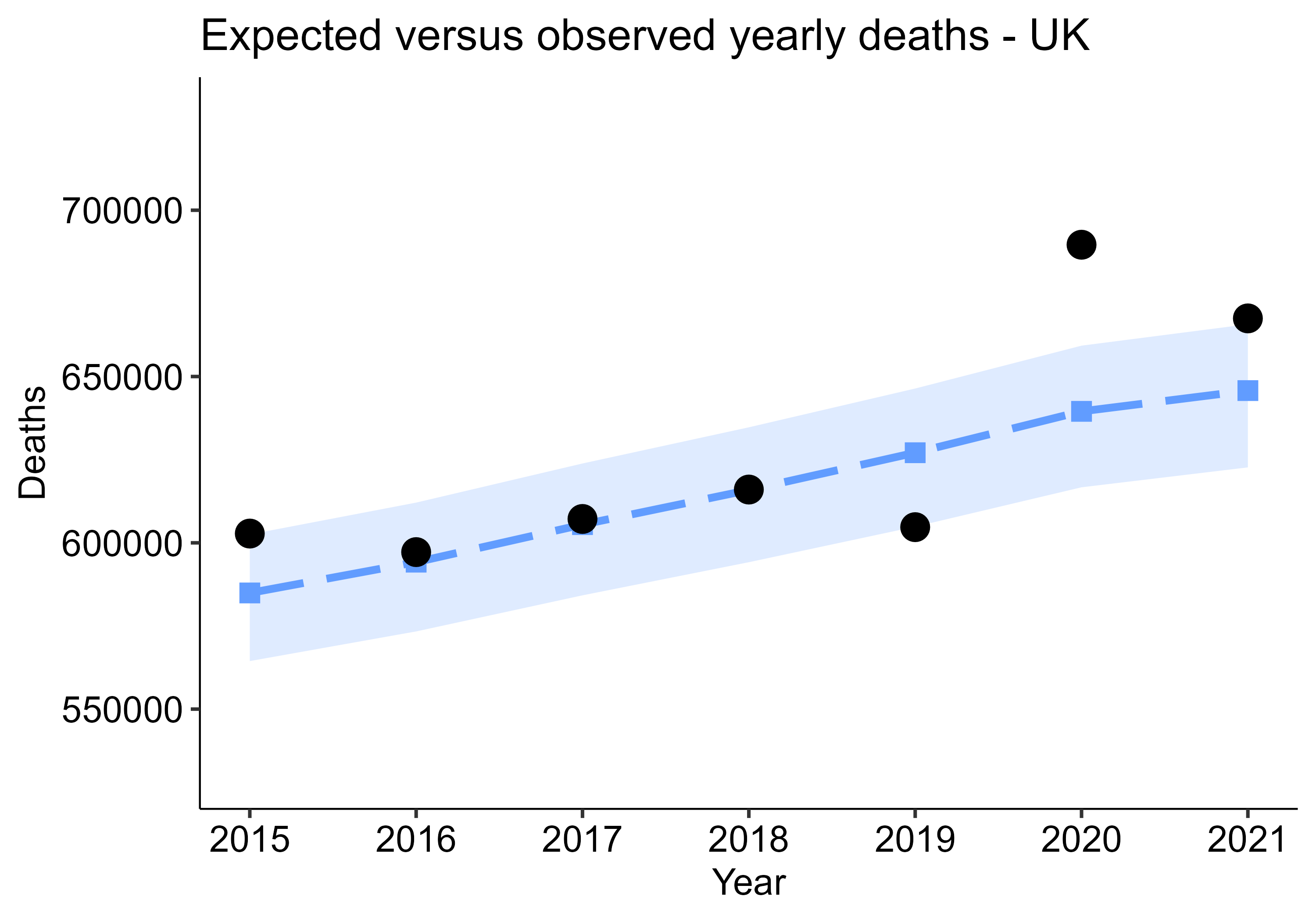} \\
\includegraphics[width = 0.496 \textwidth]{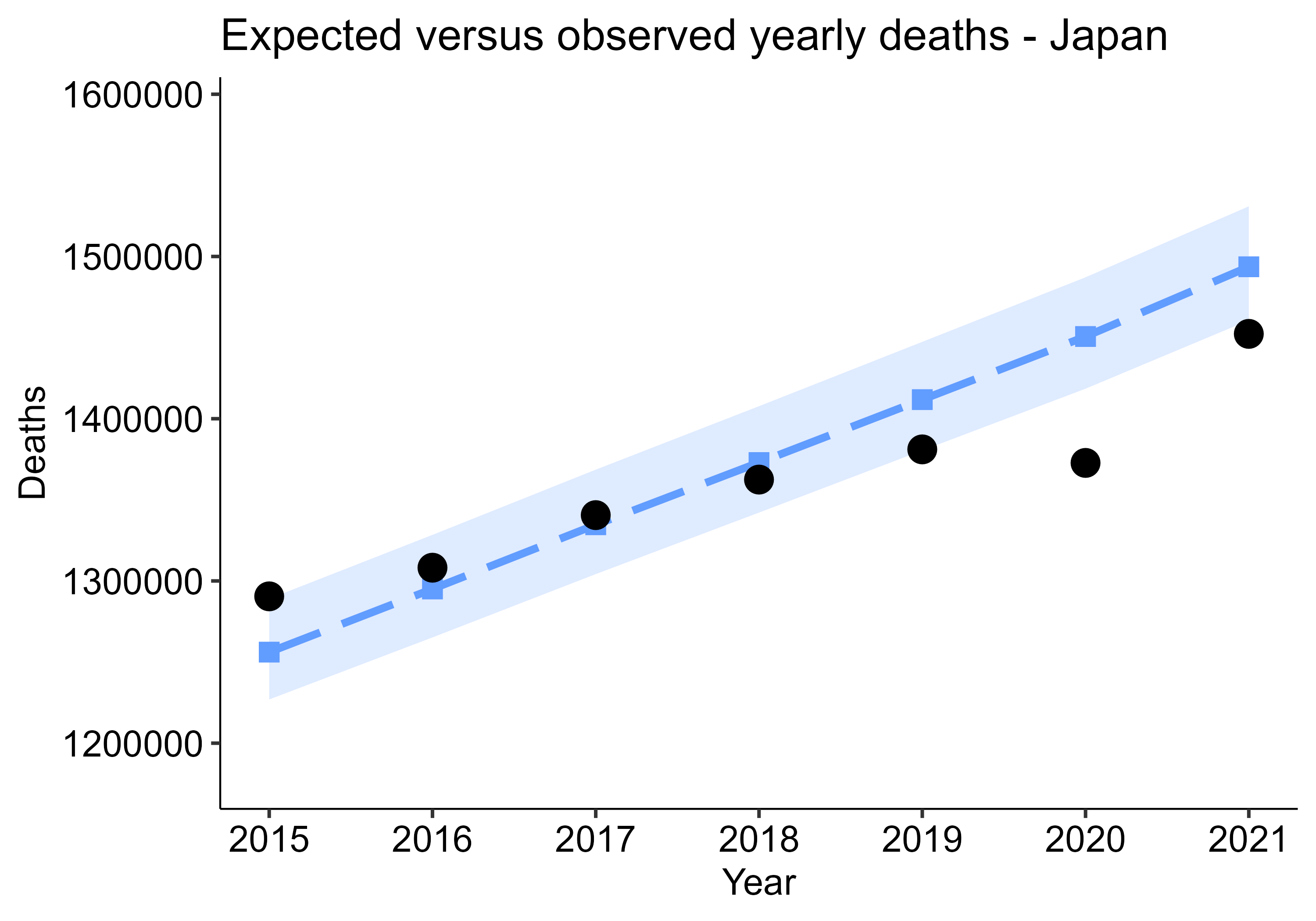} 
\includegraphics[width = 0.496 \textwidth]{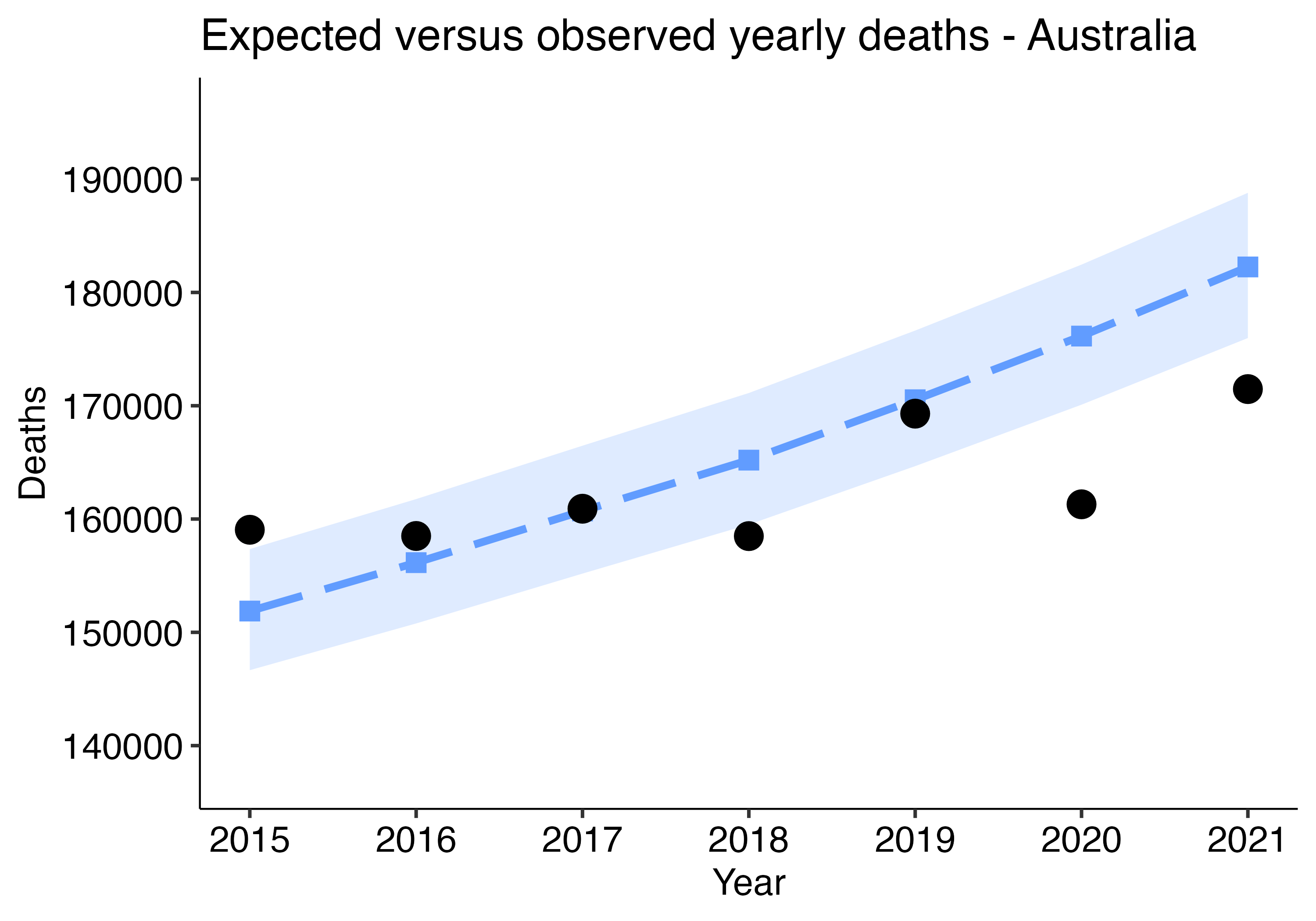} 

\caption {Expected and observed mortality figures by calendar year for four large non-EU high-income countries: USA, UK, Japan, and Australia. Black dots indicate observed mortality in a given calendar year, while blue squares indicate estimated yearly expected mortality. Shaded bands represent the estimated expected mortality range.
}
\label{fig:non-EU}
\end{figure}

Figure \ref{fig:non-EU} shows the same plot for four of the largest non-EU countries included in our analysis, namely the USA (top-left), the UK (top-right), Japan (bottom-left) and Australia (bottom-right). From the USA plot we can immediately notice the anomalous mortality levels that characterised the years 2020 and 2021 - quantifiable in 14.5\% and 16.6\% excess mortality by year, respectively. Overall, total excess deaths in the country for those two years amount to almost one million - by far the largest figure within our sample. The deaths are mainly attributable to COVID-19, with outcomes worsened by lower vaccination uptake \citep{suthar2022public} as well as conditions that may have resulted from delayed medical care and overwhelmed health systems \citep{woolf2021excess}. From the US plot we can also see how mortality from COVID-19 is especially high among the elderly: after regularly increasing from 2015 to 2020, expected mortality remains almost constant from 2020 to 2021, as the victims of COVID-19 in 2020 were in large part pertaining to the elderly population, which disproportionately contribute to expected mortality. This change in growth rate in expected mortality is also visible in other countries, such as Italy and Spain in Figure \ref{fig:EU}, as well as the UK in the top-right panel of Figure \ref{fig:non-EU}. From the latter plot, we can see how the UK also suffered from increased mortality levels in both pandemic years, even though to a lesser extent than the USA. On the other hand, the countries depicted in the bottom panels of the same figure, i.e.\ Japan and Australia, paint a completely different picture. Both nations withstood the first two pandemic years without incurring in excess mortality, and instead registered considerable mortality deficits in both years. Note that a mortality deficit during the pandemic years does not imply overreporting of COVID-19 deaths, but rather that deaths avoided or postponed by NPIs and behavioural changes in the population outweighed deaths caused by the virus. These changes may have led to, e.g.\, reduced mortality from other respiratory infections as well as accidents \citep{olsen2020decrease, barnes2022covid}. Geographical isolation of the two island countries may also have contributed to reduce the spread of COVID-19, as in the case of New Zealand \citep{kung2021reduced}. From the data in Table \ref{tab:main}, we can see how similar levels of reduced mortality during the pandemic were observed, among others, in other East Asian regions (South Korea, Taiwan, Hong Kong) as well as other islands (Iceland and New Zealand), hinting at the presence of geographical patterns. 
Plots similar to those in Figures \ref{fig:EU} and \ref{fig:non-EU} for all other countries included in our analysis are provided in the Supplementary Material (Section S.5).

\section{Comparison with other estimates}
\label{sec:comparison}

\begin{figure}[]
	\centering
 \includegraphics[width = \textwidth]{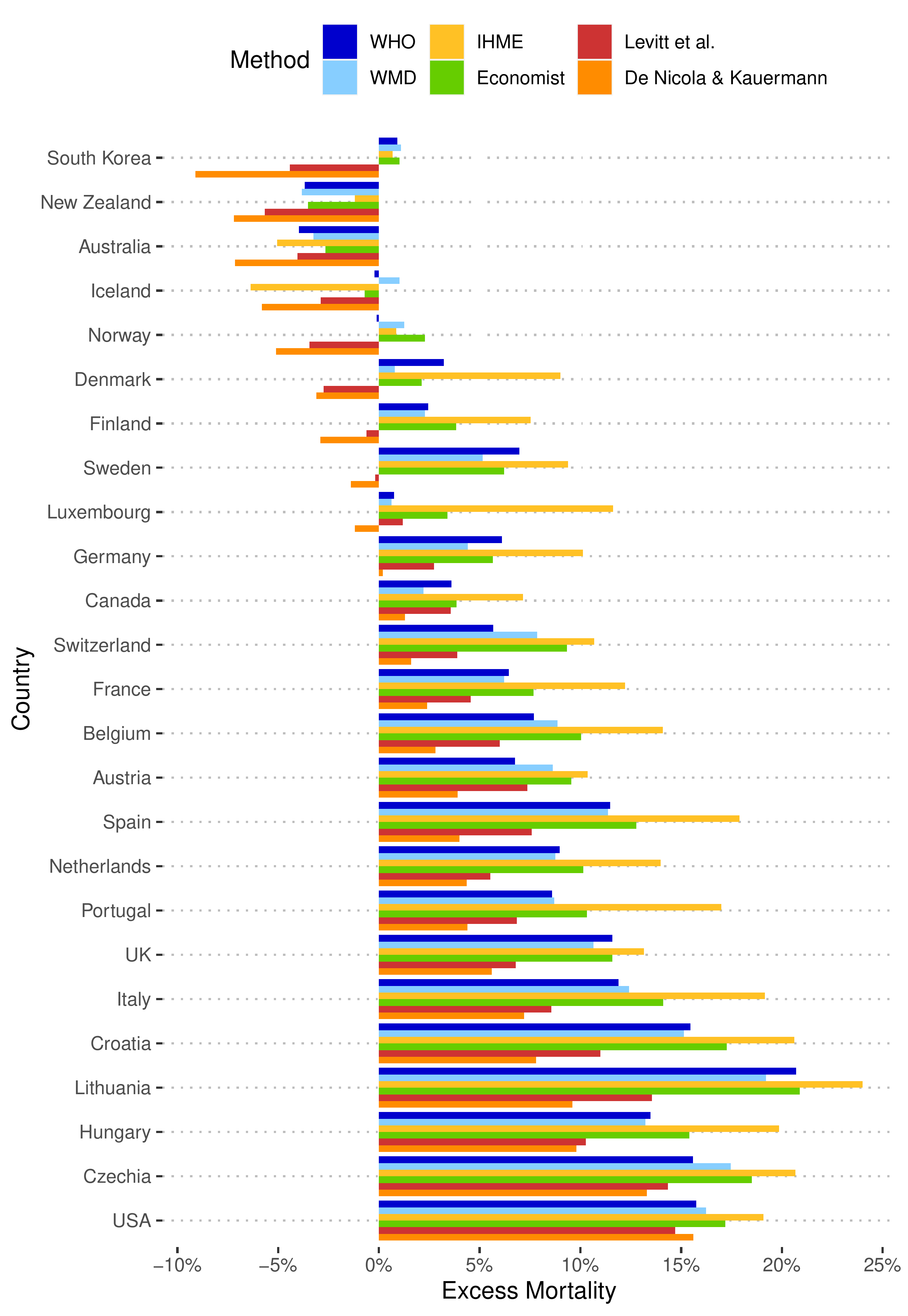}

\caption {Comparison of country-specific excess mortality estimates obtained by six different studies, including ours. Methods not accounting for age are considerably overestimating excess mortality in countries with ageing populations.}
\label{fig:comparison}
\end{figure}

This section is dedicated to comparing excess mortality estimates obtained through our method with figures produced by five other prominent multi-country studies. In particular, we contrast our results with those obtained in the already mentioned studies by \cite{msemburi2022estimates}, \cite{karlinsky2021elife}, \cite{wang2022estimating}, \cite{Economist:2023}, and \cite{levitt2022comparison}. The reasoning behind the choice of these benchmarks is the following: All five studies are very high profile, and the first four were used by prominent institutions and media outlets as their official estimates. Respectively, the first one is used by the WHO \citep{WHO:2022}, the second one for the World Mortality Database (WMD) and Our World In Data \citep{OWID:2023}, the third one by the IHME \citep{IHME:2022}, and the fourth one by The Economist. As for the fifth method by Levitt et al., while it is not currently used by official sources, we include it as we want at least one method performing some type of explicit age adjustment, and because, not unlike the other methods, it is highly published and well cited. Note that, while our analysis encompasses 30 countries, countries for which estimates are not available for all six methods were excluded from the comparison. The results for the remaining 25 countries are shown in Figure \ref{fig:comparison}, which depicts percentage excess mortality estimates for the period 2020-2021 calculated with the six different methods. Countries are shown in ascending order with respect to excess mortality computed with our method. Note that, as before, percentage excess mortality was calculated using expected mortality as the base, i.e.\ $\Delta_{t}^{\%}= {\Delta_{t}}/{E_{t}}$, and that our own measure of expected mortality was used as base for all six methods, for reasons of comparability and to ensure consistent rankings. Also note that, while results are here only presented graphically, tables containing figures for all six different methods are provided in the Supplementary Material (Section S.2).

From the figure, we can appreciate how different estimation methods result in sizeable differences in the estimates. In particular, several patterns emerge. Firstly, we can see how the IHME and Economist methods, which don't account for age at all, consistently produce the highest excess mortality estimates among the six methods. This is to be expected, due to the fact that all populations considered are ageing to at least some extent. Not considering age thus causes upward bias in the estimates. This difference is also directly related to the extent to which the population is ageing, as it is smaller and almost negligible for countries with relatively stable age pyramids, such as, e.g., the USA, while it gets larger for rapidly ageing countries, like e.g.\ Germany and France. From the plot we also see how the WHO and WMD methods tend to produce estimates that are lower than those of IHME and Economist, but still considerably larger than those of the two explicitly age-adjusted methods considered. As already mentioned in the Introduction, the WMD calculates expected mortality through use of a linear trend, and thus is only able to partially capture the effect of age, as age pyramids are generally not smooth. As for the WHO method, it uses Poisson-based spline regression to fit a smooth trend to the reference period, and then extrapolates it to calculate expected mortality in the period of interest. While trend extrapolation  can already be problematic when using linear models, splines have the additional issue of reacting strongly to short-term fluctuations. As 2019 was a generally a year of low mortality in many of the analysed countries, as visible, e.g., in Figure \ref{fig:EU}, it is likely for that to have led to overestimation in excess mortality. Shifting our focus to the remaining two methods, i.e.\ that of Levitt et al.\ and ours, we can see that the corresponding estimates not only tend to be lower than the other methods, but are also closer to each other than to the other ones. Furthermore, the two methods also rank countries similarly. %This is to be expected, given that both methods explicitly take age into account. 
However, we can also observe considerable differences between the two, with the Levitt method tending to produce slightly higher excess mortality estimates. These differences are, in some cases, in the order of 5\%. We believe this to be mainly due to two factors. Firstly, the Levitt method uses 2017-2019 as  reference period, while we use 2015-2019. Given that mortality was overall slightly higher in 2015 and 2016 than in the 2017-2019 window, estimates of expected mortality based only on the latter years will naturally be slightly lower. In our opinion, both choices of reference period are equally valid, with each having pros and cons. Crucially, our choice of a five-years reference period allows us to more effectively apply our uncertainty quantification approach by using yearly age-specific mortality rates. Nonetheless, we also recalculated our mean estimates using the same 2017-2019 window as the reference period, and included the results in the Supplementary Material (Section S.3). The alternative reference period results in slightly higher excess mortality estimates, while leaving country rankings largely unaffected (see also \citealp{levitt2023excess}). More importantly, differences between the two sets of estimates persist even after aligning the reference periods, leading us to discuss the second main distinguishing factor between the two methods, namely the method used to perform age adjustment. Levitt et al.'s procedure consists in dividing the population into five age strata (0-14, 15-64, 65-74, 75-85, and >85 years old) and calculating raw excess mortality for each strata, before summing them up to obtain the overall figure. While this is far better than no age adjustment at all, it still disregards age variation within the age groups. This can be especially problematic if the age classes are large, as for example the central age class of 15-64 used by the authors, or if variation within the age classes is large, as e.g.\ in the 75-85 group for of Germany, depicted in Figure \ref{fig:pyramids}. %This reduces the magnitude of the age-induced bias, and is thus certainly better than not accounting for age at all; on the other hand, simply partitioning the population into age groups is equivalent to assuming age structure to be homogeneous within those age groups, which is unrealistic for, e.g.\, the age group 15-64 in ageing European countries, where the older part of the group (i.e. close to 64 years of age) is decisively more numerous than the younger part (i.e.\ close to 15 years), as visible in Figure \ref{fig:pyramids} for the case of Germany. 
The latter point may thus constitute a cause of bias in excess mortality estimation. In contrast, our method performs age standardisation by partitioning the  population pyramid at the finest available resolution of the data, i.e.\ 1-year age classes, which we argue to be preferable when data to do so is available.

%The best solution seems to be to directly do so, as done by the Levitt method and our method. Still, these two methods produce significantly different results from one another. In particular, our method almost always produces lower excess mortality estimates. This is to be expected given that the Levitt method standardizes using large age classes, therefore assuming those age classes to be substantially homogeneous. But we know that this is not the case in ageing high income countries, where, e.g.\ , the age class 15-64 contains substantially more people closer to age 64 than 14. Interestingly, this is not the case for the USA, the only wealthy country in which population ageing is not currently an issue. This is most likely the reason why our method actually estimates a higher excess mortality for the USA as compared to Levitt. The somewhat stable age structure of the United States is also why our excess mortality estimate for the country does not differ dramatically from all other methods, as age adjustment does not have a big impact.  

\section{Discussion}
\label{sec:discussion}

Accurately measuring excess deaths in times of COVID-19 is vital to assess the pandemic's impact on public health across different countries and regions of the world. Precise excess mortality estimation is also crucial for explaining the pandemic curve, for understanding the factors contributing to differences in the infection-fatality rate among populations, and for gauging the effectiveness of alternative policy options for managing future crises. In this paper, we demonstrated the importance of taking age into account to obtain unbiased excess mortality estimates in high-income countries, and proposed a simple and effective method to do so when the necessary data is available. We applied our method to 30 different countries, and compared our results with other estimates obtained by five other high profile studies. The comparison sheds light on how different estimation methods result in sizeable differences in the estimates. Results are especially sensitive to the absence of age adjustment, which can lead to considerable downward bias, and is most relevant for countries with rapidly ageing populations. For example, estimates in South Korea, one of the countries with the lowest total fertility rates in the world, range from slightly positive excess mortality for methods without age adjustment to a 9.2\% mortality deficit using our method. Large differences are also observable in most other high-income countries, while estimates for populations ageing at a slower rate, such as, e.g., the USA, are more convergent. 
It is thus crucial to explicitly take age into account for excess mortality estimation in ageing high-income countries. Note that the focus of most of the studies included in the comparison does not lie specifically on high-income countries, but rather on obtaining global estimates of excess deaths, including situations in which high-quality data is not available. Nonetheless, even in the case of global estimates, we argue that it would be relevant to incorporate age adjustment in the estimates for countries in which data to do so is available, as the latter make up a non-negligible portion of the global population. Moreover, we are fortunate that countries where age adjustment tends to be most impactful are also the ones for which high-resolution data is available, making unbiased estimation possible. From the comparison, we further see that performing age adjustment at a fine level, as opposed to doing so by dividing the population into large age classes (as done by \citealp{levitt2022comparison}), also results in considerable differences in excess mortality estimates. Unsurprisingly, these differences are also more pronounced for countries in which the structure of the population pyramid is less stable over time. This speaks to the importance of utilising high-resolution age-stratified data to perform the estimation.

Turning our attention from the method to the empirical results obtained, our estimates uncover large variation in pandemic outcomes among the 30 analysed countries. More specifically, over the 2020-2021 period, 10 countries showed excess mortality beyond what could be explained by standard variation, and 8 displayed a sizeable mortality deficit. In the remaining 12 countries mortality was neither markedly higher nor lower than during the 2015-2019 reference period. The countries with the worst outcomes relative to population were Bulgaria, the only non-high-income country in our sample, which had a 2-years excess of 22.8\%, and the United States, which have seen a 15.6\% increase in mortality over the pandemic period. The latter increase is particularly notable given the size of the country's population, as it corresponds to more than 920.000 excess deaths over the two years. Other countries that experienced excess mortality (in the order of 5 to 10\%)  include most of the Eastern and Southern European states, as well as the UK. In contrast, considerable mortality deficits of similar magnitude were observed in some of the Nordic European countries, as well as in Australia, New Zealand, South Korea, Taiwan and Japan. We here want to stress that, while the absence of excess mortality (or the presence of a mortality deficit) in a given country does indicate that the country fared well in terms of life loss during the pandemic, it does not imply that no people died because of COVID-19. Instead, reduced mortality during pandemic years simply indicates that deaths from other causes which were prevented during the pandemic, e.g.\ through governmental NPIs and/or behavioural changes in the population, outnumbered COVID-19 victims. Likewise, for the same reasons, a value of overall excess mortality lower than the officially reported COVID-19 death toll in a given country does not imply overreporting of COVID-related deaths. In fact, if the healthcare system keeps functioning normally (i.e.\ the same way as before the pandemic), one would actually expect there to be less overall excess mortality than total deaths caused by COVID-19, as lockdowns can prevent or postpone deaths from other causes, such as e.g.\ accidents or  other respiratory infections \citep{calderon2021impact,olsen2020decrease}. Assuming a perfectly functioning reporting system, this should also be reflected in the official death tolls (note that official COVID-19 death tolls for the studied period are given in the Supplementary Material, for comparison purposes). Of course, even within our high-income sample, the latter assumption only holds to a certain degree in practice, as reporting systems can vary substantially between countries \citep{karanikolos2020comparable}. Likewise, the assumption on the functioning healthcare system is not always valid, as overwhelmed healthcare systems were reported at various stages of the crisis \citep{senni2020bergamo,dorsett2020point,alderwick2022nhs}. Issues with reporting and healthcare systems pose even bigger problems in countries falling within low- and middle-income brackets \citep{chatterjee2020india, lone2020covid}. It is therefore not a random occurrence that Bulgaria, being the sole non-high-income country in our sample, displays the highest excess mortality of all analyzed regions. Indeed, our selection of countries is not at all representative of the whole world, and outcomes are likely to be much worse in regions with fragile and less efficient healthcare systems \citep{bong2020covid}. To obtain global estimates one would therefore need to use estimation techniques which work with deficient data, which our method does not do. Further, we note that our method performs retrospective excess mortality estimation using historical data, which is useful to understand the impact of a crisis and to learn policy lessons in preparation for future ones. On the other hand, the need to use full historical data limits its effectiveness for managing an ongoing crisis, for which a real-time monitoring tool would be needed. While it would be possible to use weekly data as soon as they are available, as done e.g.\ in Section 3 of \cite{de2022assessing}, complete weekly data still usually arrives with a delay of at least several weeks, rendering it unfit for live monitoring. A possibility for developing a truly real-time tool would be to make use of so-called nowcasting techniques, i.e.\ methods bridging the delay between events and their reporting, to estimate the number of fatalities which already occurred based on the ones that were already reported. Examples of this are given by \cite{schneble2021nowcasting} and \cite{de2022regional}, who perform nowcasting for fatal and general COVID-19 infections, respectively. While this is beyond the scope of this paper, nowcasting all-cause mortality data to create a real-time monitoring tool is certainly an interesting direction for future research, with great potential for real-world impact. Another thing to note is that our method for calculating expected mortality does not account for potential trends in life expectancy over time. This is equivalent to implicitly assuming constant age-specific hazards over the considered years (while we still, of course, account for the evolving age structure of the population). As discussed in the introduction, using a time trend to project changes in death rates observed during the reference period on the period of interest can lead to instability in the estimates, due to the large degree of natural variation that is present in all-cause deaths. Furthermore, there is no guarantee that mortality rates should continue to follow the same trend that was observed during the reference period, even in the absence of major perturbation events \citep{levitt2023excess, ioannidis2023flaws}. For these reasons, we opted to not incorporate any trend, and instead simply use the age-specific average death rates over the 5-years reference period. %as discussed in intro, incorporating pure trend can lead to issues 1). Further, 2). Noentheless, if projecting on longer period one should try with e.g. modified version of lee-carter mitigating variations, e.g. incorporating a cross-country trend instead of country-specific ones.
This work particularly well in our case, as both the period of interest and the reference period are relatively short, and changes in life expectancy over the reference period were generally moderate in high-income countries \citep{aburto2022quantifying}. However, if one would aim at estimating excess mortality over a longer period following the pandemic, accounting for (expected) changes in life expectancy would be recommendable. While certainly not straightforward, as it requires several further assumptions, such an adjustment could be attempted by adapting projection techniques (see e.g. \citealp{lee2000lee}) while still keeping the estimators' variance in check by, e.g., incorporating cross-country time trends instead of country-specific ones.

One of the features of our method is that, in addition to point estimates, it also allows to produce excess mortality ranges, providing us with best- and worst-case mortality scenarios under conditions observed during the reference period. We stress that these are not classical confidence intervals, and as such they do not give us a probabilistic measure of uncertainty. As detailed in Section \ref{sec:methods}, however, calculating standard confidence intervals would require imposing unconvincing distributional assumptions on both the reference population and the mortality process, thus injecting a large amount of model-related uncertainty in the figures. We, therefore, opted for the data-driven, multiverse-style approach described, which is considerably more robust and allows us to make clear statements on whether or not mortality was substantially different in the period of interest than in the reference period. 

Given the primary role of age in both overall and COVID-related mortality, incorporating it into the estimation in some way is essential to obtain unbiased estimates. With this paper, we hope to make explicit age adjustment a standard practice for excess mortality estimation in cases for which age-stratified data is available. To this avail, we have publicly shared all data and code relevant to this study, to facilitate researchers in reproducing our findings as well as to enable them to utilise and build on our methods for future applications.

\section*{Data and code availability}
We provide full replication code and materials, including data, in our GitHub repository, available at \href{https://github.com/gdenicola/excess-mortality-world}{\texttt{https://github.com/gdenicola/excess-mortality-world}}. All data used was gathered from publicly available sources, with details on the specific sources given in Section \ref{sec:data} of the manuscript, the Supplementary Material, and the repository itself.

\section*{Acknowledgments}
The authors would like to thank Juan Camilo Rosas Romero as well as all members of the COVID-19 Data Analysis Group (CODAG@LMU) for invaluable comments and fruitful discussions. 

% With this work, we hope to establish a standard practice for estimating expected mortality in countries where fine-grained age-specific data is available. To this avail, we also make all data and code necessary to reproduce the analysis available in our public GitHub repository \citep{denicola_repository}.
% %\footnote{\url{https://github.com/gdenicola/excess-mortality-world}}. 
% This is done to ease the reproduction of our results, and to facilitate researchers in employing and adapting our methods for different applications. Above all, our analysis demonstrates the importance of proper age adjustment in producing unbiased estimates for all-cause excess mortality.

\bibliography{literature.bib}

\newpage
%\appendix
%\section{}
%\label{sec:annex}
%TC:ignore

\renewcommand{\thefootnote}{\fnsymbol{footnote}} %THIS IS THE FOOTNOTE COMMAND

\begin{center}
    %\textbf{Modelling large and dynamically growing bipartite networks - A case study in patent data \\}
	{\Large{{Supplementary Material}} \\
 \medskip
 \Large{\textbf{Estimating excess mortality in high-income countries during the COVID-19 pandemic}}}  \\ 
	Giacomo De Nicola$^{1,}$\footnote[1]{Corresponding author:  giacomo.denicola@stat.uni-muenchen.de}, Göran Kauermann$^{1}$\hspace{.2cm}\\
	\vspace{0.3cm}
	Department of Statistics, LMU Munich$^1$ \\
\end{center}

\setcounter{section}{0}
\setcounter{page}{1}

\makeatletter 
\setcounter{figure}{0}
\setcounter{table}{0}
\renewcommand{\thefigure}{S.\@arabic\c@figure}
\renewcommand{\thetable}{S.\@arabic\c@table}
\renewcommand{\thesection}{S.\@arabic\c@section}

\makeatother

\setcounter{equation}{0}

\noindent{The following contains further results omitted from the paper for reasons of brevity, and provides additional depth and detail to the manuscript. The full data and code to reproduce the analysis can be found in our {Github repository}, available at: }

%\vspace{0.1cm}
\begin{center}
    
{\href{https://github.com/gdenicola/excess-mortality-world}{\texttt{https://github.com/gdenicola/excess-mortality-world}}. }

\end{center}
%\vspace{0.1cm}

\noindent{The repository also contains additional information on the data sources used for the analyses, as well as direct links to each of them}.

\section{Year-specific excess mortality}

Tables \ref{tab:2020} and \ref{tab:2021} contain year-specific excess mortality estimates for 2020 and 2021, respectively. More specifically, similarly as in Table \ref{tab:main} in the main body of the paper, expected, observed and excess mortality figures are given. The tables further provide percentage excess mortality estimates as well as plausible ranges for each of the two years, calculated as detailed in Section \ref{sec:methods} of the paper.

\begin{table}[ht]
	\centering
	\begin{tabular}{lrrrrr}
		\hline
		Country & Population & Expected & Observed &  Excess  & \%Excess  \\
		\hline
		Australia & 25,496,892 & 176,148  & 161,300  & -14,848 & -8.4\% \\ 
		Austria & 8,901,056 & 87,861 & 91,559 &  3,738 & +4.3\% \\ 
		Belgium & 11,522,436 & 116,190 & 126,896 & 10,706  & +9.2\% \\ 
		Bulgaria & 6,951,471 & 111,477 &  124,735 & 13,258 & +11.9\% \\ 
		Canada & 37,777,751 & 301,511 & 308,412 & 6,901 & +2.3\% \\ 
		Croatia & 4,058,161 & 55,338 & 57,023 & 1,685 & +3.0\% \\ 
		Czechia & 10,555,979 & 118,246 & 129,289 & 11,043 & +9.3\% \\ 
		Denmark & 5,822,777 & 56,991 & 54,645 & -2,346 & -4.1\% \\ 
  		Finland & 5,525,274 & 57,565  &  55,488 & -2,077 & -3.6\% \\ 
		France & 65,287,143 & 630,073  & 654,599 & 24,526 & +3.9\% \\ 
		Germany & 83,165,450 & 993,863 & 985,572 & -8,291  & -0.8\% \\ 
		Hong Kong & 7,494,450 & 53,193 & 50,653 & -2,540 & -4.8\% \\ 
		Hungary & 9,769,852 & 134,947 & 141,326 & 6,379 & +4.7\% \\ 
		Iceland & 364,132 & 2,428 & 2,304 & -124 & -5.1\% \\ 
		Ireland & 4,962,883 & 33,663 & 32,387 & -1,276 & -3.8\% \\ 
		Italy & 59,641,488 & 674,859 & 740,317 & 65,458 & +9.7\% \\ 
  		Japan & 123,735,685 & 1,450,666  & 1,372,755 & -77,911 & -5.4\% \\ 
		Lithuania & 2,794,066 & 41,542  &  43,547 & 2,005 & +4.8\% \\ 
		Luxembourg & 626,059 & 4,559 & 4,609 & 50  & +1.1\% \\ 
  	Netherlands & 17,407,606 & 161,349 & 168,678 & 7329 & +4.5\% \\ 
		New Zealand & 5,034,795 & 35,717 & 32,613 & -3,104 & -8.7\% \\ 
		Norway & 5,367,573 & 43,032 & 40,611 & -2,421 & -5.6\% \\ 
		Portugal & 10,315,540 & 118,173 & 123,396 & 5,223 & +4.4\% \\ 
		South Korea & 51,348,579 & 333,713 & 304,948 & -28,765 & -8.6\% \\ 
		Spain & 47,330,878 & 450,585 & 492,447 & 41,862 & +9.3\% \\ 
            Sweden & 10,327,450 & 95,804 & 98,124 & 2,320 & +2.4\% \\ 
            Switzerland & 8,605,887 & 72,060 & 76,195 & 4,135 & +5.7\% \\ 
            Taiwan & 23,600,867 & 188,650 & 173,067 & -15,583 & -8.3\% \\ 
            UK & 66,939,020 & 639,528 & 689,629 & 50,101 & +7.8\% \\ 
            USA & 329,791,078 & 2,955,815 & 3,383,729 & 427,914 & +14.5\% \\ 
		\hline
	\end{tabular}
	\caption{Expected and observed yearly mortality in 2020 for each of the 30 countries included in the analysis.}
	
	\label{tab:2020}
\end{table}

\begin{table}[ht]
	\centering
	\begin{tabular}{lrrrrr}
		\hline
		Country & Population & Expected & Observed & Excess  &  \%Excess \\ 
		\hline
		Australia & 25,671,819 & 182,249  &  171,469 & -10,780 & -5.9\% \\ 
		Austria & 8,932,664 & 88,875  & 91,962 & 3,087 & +3.5\% \\ 
		Belgium & 11,554,787 & 116,567 & 112,331 & -4,236  & -3.6\% \\ 
		Bulgaria & 6,916,530 & 111,422 &  148,995 & 37,572 & +33.7\% \\ 
		Canada & 38,099,361 & 309,631 & 311,640 & 2,009 & +0.6\% \\ 
		Croatia & 4,036,355 & 55,755 & 62,712 & 6,957 & +12.5\% \\ 
		Czechia & 10,541,134 & 119,242 & 139,891 & +20,649 & +13.3\% \\ 
		Denmark & 5,840,049 & 58,357 & 57,152 & -1,205 & -3.1\% \\ 
  		Finland & 5,533,778 & 58,919  &  57,659 & -1,260 & -2.1\% \\ 
		France & 65,447,454 &  637,968 & 644,201 & 6,233 & +1\% \\ 
		Germany & 83,155,031 & 1,011,298 & 1,023,687 & 12,389 & +1.2\% \\ 
		Hong Kong & 7,486,815 & 55,435 & 51,536 & -3,899 & -7.0\% \\ 
		Hungary & 9,730,772 & 135,857 & 156,131 & 20,274 & +14.9\% \\ 
		Iceland & 368,792 & 2,498 & 2,333 & -165 & -6.6\% \\ 
		Ireland & 5,001,674 & 34,741 & 33,058 & -1,683 & -4.8\% \\ 
		Italy & 59,236,213 & 677,602 & 709,035 & 31,433 & +4.6\% \\ 
  		Japan & 123,244,767 & 1,493,644  & 1,452,289 & -41,335 & -2.8\% \\ 
		Lithuania & 2,795,680 & 41,750  &  47,746 &  5,995 & +14,4\% \\ 
		Luxembourg & 634,703 & 4,652 & 4,489 & -163  & -3.5\% \\ 
  		Netherlands & 17,475,445 & 164,126 & 170,972 & 6,846 & +4,2\% \\ 
		New Zealand & 5,101,593 & 37,030 & 34,932 & -2,098 & -5.7\% \\ 
		Norway & 5,391,369 & 43,999 & 42,002 & -1,997 & -4.5\% \\ 
		Portugal & 10,315,101 & 119,572 & 124,802 & 5,230 & +4.4\% \\ 
		South Korea & 51,349,116 & 350,949 & 317,680 & -33,269 & -9.5\% \\ 
		Spain & 47,398,697 & 454,822 & 449,270 & -5,552 & -1.2\% \\ 
            Sweden & 10,379,247 & 96,959 & 91,958 & -5,001 & -5.2\% \\ 
            Switzerland & 8,670,243 & 73,071 & 71,192 & -1,879 & -2.6\% \\ 
            Taiwan & 23,560,033 & 194,821 & 184,172 & -10,649 & -5.5\% \\ 
            UK & 66,930,951 & 645,772 & 667,479 & 21,707 & +3.4\% \\ 
            USA & 331,697,950 & 2,965,880 & 3,458,697 & 492,817 & +16.6\% \\ 
		\hline
	\end{tabular}
	\caption{Expected and observed yearly mortality in 2021 for each of the 30 countries included in the analysis.}
	
	\label{tab:2021}
\end{table}

\section{Comparison tables}
\label{sec:comptab}

This section contains excess mortality figures computed with each of the seven different methods discussed in the manuscript (i.e.\ our method and the five methods with which the comparison is carried out). More specifically, Table \ref{tab:comparison_absolute} contains absolute excess mortality figures for all methods, while Table \ref{tab:comparison_relative} gives the same figures on a relative basis. The latter table was used to produce Figure \ref{fig:comparison} in Section \ref{sec:comparison} of the manuscript. Note that, as in the main body of the paper, percentage excess mortality was calculated using expected mortality as the base, i.e.\ $\Delta_{t}^{\%}= {\Delta_{t}}/{E_{t}}$, and that our own measure of expected mortality was used as base for all six methods, for reasons of comparability and consistency. Finally, table \ref{tab:comparison_rates} contains excess deaths per million inhabitants by country for the period 2020-2021, estimated through the six different methods compared in the paper.

\begin{table}[!ht]
    \centering
    \begin{tabular}{lrrrrrrr}
    \hline
        Country & Observed & WMD & Economist & IHME & WHO & Levitt & De Nicola  \\ 
        \hline
        Australia & 334,035 & -11,639 & -9,500 & -18,100 & -14,255 & -14,460 & -25,628  \\ 
        Austria & 183,561 & 15,261 & 16,877 & 18,300 & 11,938 & 13,007 & 6,825  \\ 
        Belgium & 239,227 & 20,613 & 23,364 & 32,800 & 17,918 & 13,958 & 6,470  \\ 
        Canada & 618,349 & 13,474 & 23,548 & 43,700 & 22,019 & 21,829 & 8,910  \\ 
        Croatia & 119,735 & 16,826 & 19,186 & 22,900 & 17,176 & 12,205 & 8,642  \\ 
        Czechia & 269,180 & 41,480 & 43,942 & 49,100 & 37,039 & 34,079 & 31,691  \\ 
        Denmark & 111,797 & 913 & 2,453 & 10,400 & 3,716 & -3,157 & -3,552  \\ 
        Finland & 113,147 & 2,662 & 4,469 & 8,780 & 2,857 & -716 & -3,337  \\ 
        France & 1,298,800 & 78,910 & 97,390 & 155,000 & 81,850 & 57,767 & 30,759  \\ 
        Germany & 2,009,259 & 88,446 & 113,242 & 203,000 & 122,432 & 54,740 & 4,098  \\ 
        Hungary & 297,457 & 35,811 & 41,714 & 53,800 & 36,497 & 27,813 & 26,653  \\ 
        Iceland & 4,637 & 50 & -35 & -314 & -11 & -142 & -289  \\ 
        Italy & 1,449,352 & 167,816 & 190,872 & 259,000 & 160,801 & 115,690 & 96,891  \\ 
        Lithuania & 91,293 & 16,008 & 17,396 & 20,000 & 17,255 & 11,283 & 8,000  \\ 
        Luxembourg & 9,098 & 57 & 314 & 1,070 & 69 & 109 & -113  \\ 
        Netherlands & 339,650 & 28,495 & 33,017 & 45,500 & 29,213 & 17,969 & 14,175  \\ 
        New Zealand & 67,545 & -2,787 & -2,566 & -872 & -2,677 & -4,118 & -5,201  \\ 
        Norway & 82,613 & 1,101 & 1,986 & 742 & -101 & -2,994 & -4,418  \\ 
        Portugal & 248,198 & 20,677 & 24,530 & 40,400 & 20,447 & 16,286 & 10,454  \\ 
        South Korea & 622,628 & 7,529 & 6,967 & 4,630 & 6,288 & -30,286 & -62,034  \\ 
        Spain & 941,717 & 102,991 & 115,685 & 162,000 & 103,937 & 68,720 & 36,310  \\ 
        Sweden & 190,082 & 9,926 & 11,976 & 18,100 & 13,439 & -367 & -2,681  \\ 
        Switzerland & 147,387 & 11,394 & 13,539 & 15,500 & 8,247 & 5,640 & 2,256  \\ 
        UK & 1,357,108 & 136,795 & 148,889 & 169,000 & 148,897 & 87,307 & 71,808  \\ 
        USA & 6,842,426 & 961,032 & 1,017,655 & 1,130,000 & 932,458 & 871,295 & 920,731  \\
        \hline
    \end{tabular}
    \caption{Absolute excess deaths for the period 2020-2021, estimated through the six different methods compared in the paper.}
        \label{tab:comparison_absolute}
\end{table}

\begin{table}[!ht]
    \centering
    \begin{tabular}{lrrrrrr}
    \hline
        Country & WMD\% & Economist \% & IHME \% & WHO \% & Levitt \% & De Nicola \%  \\ 
        \hline
        Australia & -3.2\% & -2.7\% & -5.1\% & -4.0\% & -4.0\% & -7.2\%  \\ 
        Austria & 8.6\% & 9.5\% & 10.4\% & 6.8\% & 7.4\% & 3.9\%  \\ 
        Belgium & 8.9\% & 10.0\% & 14.1\% & 7.7\% & 6.0\% & 2.8\%  \\ 
        Canada & 2.2\% & 3.9\% & 7.2\% & 3.6\% & 3.6\% & 1.5\%  \\ 
        Croatia & 15.1\% & 17.3\% & 20.6\% & 15.5\% & 11.0\% & 7.8\%  \\ 
        Czechia & 17.5\% & 18.5\% & 20.7\% & 15.6\% & 14.3\% & 13.3\%  \\ 
        Denmark & 0.8\% & 2.1\% & 9.0\% & 3.2\% & -2.7\% & -3.1\%  \\ 
        Finland & 2.3\% & 3.8\% & 7.5\% & 2.5\% & -0.6\% & -2.9\%  \\ 
        France & 6.2\% & 7.7\% & 12.2\% & 6.5\% & 4.6\% & 2.4\%  \\ 
        Germany & 4.4\% & 5.6\% & 10.1\% & 6.1\% & 2.7\% & 0.2\%  \\ 
        Hungary & 13.2\% & 15.4\% & 19.9\% & 13.5\% & 10.3\% & 9.8\%  \\ 
        Iceland & 1.0\% & -0.7\% & -6.4\% & -0.2\% & -2.9\% & -5.8\%  \\ 
        Italy & 12.4\% & 14.1\% & 19.2\% & 11.9\% & 8.6\% & 7.2\%  \\ 
        Lithuania & 19.2\% & 20.9\% & 24.0\% & 20.7\% & 13.5\% & 9.6\%  \\ 
        Luxembourg & 0.6\% & 3.4\% & 11.6\% & 0.8\% & 1.2\% & -1.2\%  \\ 
        Netherlands & 8.8\% & 10.1\% & 14.0\% & 9.0\% & 5.5\% & 4.4\%  \\
        New Zealand & -3.8\% & -3.5\% & -1.2\% & -3.7\% & -5.7\% & -7.2\%  \\
        Norway & 1.3\% & 2.3\% & 0.9\% & -0.1\% & -3.4\% & -5.1\%  \\ 
        Portugal & 8.7\% & 10.3\% & 17.0\% & 8.6\% & 6.9\% & 4.4\%  \\ 
        South Korea & 1.1\% & 1.0\% & 0.7\% & 0.9\% & -4.4\% & -9.1\%  \\ 
        Spain & 11.4\% & 12.8\% & 17.9\% & 11.5\% & 7.6\% & 4.0\%  \\ 
        Sweden & 5.1\% & 6.2\% & 9.4\% & 7.0\% & -0.2\% & -1.4\%  \\ 
        Switzerland & 7.9\% & 9.3\% & 10.7\% & 5.7\% & 3.9\% & 1.6\%  \\ 
        UK & 10.6\% & 11.6\% & 13.1\% & 11.6\% & 6.8\% & 5.6\%  \\ 
        USA & 16.2\% & 17.2\% & 19.1\% & 15.7\% & 14.7\% & 15.6\%  \\ 
        \hline
    \end{tabular}
    \caption{Relative excess deaths for the period 2020-2021, estimated through the six different methods compared in the paper. These are the percentages plotted in Figure \ref{fig:comparison}.}
    \label{tab:comparison_relative}
\end{table}

\begin{table}[!ht]
    \centering
    \begin{tabular}{lrrrrrr}
    \hline
        Country & WMD & Economist & IHME & WHO & Levitt & De Nicola \\ 
        \hline
        Australia & -227.5 & -185.7 & -353.7 & -278.6 & -282.6 & -500.9   \\
        Austria & 855.7 & 946.4 & 1026.1 & 669.4 & 729.3 & 382.7  \\ 
        Belgium & 893.2 & 1012.4 & 1421.3 & 776.4 & 604.8 & 280.4  \\
        Canada & 177.6 & 310.3 & 575.9 & 290.2 & 287.7 & 117.4  \\ 
        Croatia & 2078.7 & 2370.2 & 2829.1 & 2121.9 & 1507.8 & 1067.6  \\
        Czechia & 1966.1 & 2082.8 & 2327.3 & 1755.6 & 1615.3 & 1502.1  \\ 
        Denmark & 78.3 & 210.3 & 891.7 & 318.6 & -270.7 & -304.6  \\ 
        Finland & 240.7 & 404.1 & 793.9 & 258.3 & -64.7 & -301.7 \\ 
        France & 603.6 & 744.9 & 1185.6 & 626.1 & 441.9 & 235.3 \\ 
        Germany & 531.8 & 680.9 & 1220.5 & 736.1 & 329.1 & 24.6  \\ 
        Hungary & 1836.4 & 2139.1 & 2758.9 & 1871.6 & 1426.3 & 1366.8  \\ 
        Iceland & 68.2 & -47.8 & -428.4 & -15.0 & -193.7 & -394.3  \\ 
        Italy & 1411.7 & 1605.6 & 2178.7 & 1352.7 & 973.2 & 815.0  \\
        Lithuania & 2863.8 & 3112.1 & 3578.0 & 3086.9 & 2018.5 & 1431.2  \\
        Luxembourg & 45.2 & 249.1 & 848.7 & 54.7 & 86.5 & -89.6  \\ 
        Netherlands & 816.9 & 946.5 & 1304.4 & 837.5 & 515.1 & 406.4  \\ 
        New Zealand & -275.0 & -253.1 & -86.0 & -264.1 & -406.3 & -513.1  \\ 
        Norway & 102.3 & 184.6 & 69.0 & -9.4 & -278.3 & -410.6  \\ 
        Portugal & 1002.2 & 1189.0 & 1958.3 & 991.1 & 789.4 & 506.7  \\ 
        South Korea & 73.3 & 67.8 & 45.1 & 61.2 & -294.9 & -604.0  \\ 
        Spain & 1087.2 & 1221.2 & 1710.1 & 1097.2 & 725.4 & 383.3  \\ 
        Sweden & 479.4 & 578.4 & 874.1 & 649.0 & -17.7 & -129.5  \\
        Switzerland & 659.5 & 783.7 & 897.2 & 477.4 & 326.5 & 130.6  \\ 
        UK & 1021.8 & 1112.2 & 1262.4 & 1112.3 & 652.2 & 536.4  \\
        USA & 1452.8 & 1538.4 & 1708.3 & 1409.6 & 1317.2 & 1391.9 \\ 
        \hline

    \end{tabular}
    \caption{Yearly excess deaths per million inhabitants for the period 2020-2021, estimated through the six different methods compared in the paper.}
    \label{tab:comparison_rates}
\end{table}

\section{Estimates with 2017-2019 as reference period}
\label{sec:threeyears}

Table \ref{tab:reference} compares excess mortality figures obtained in the main paper, obtained with a 5 years reference period (2015-2019), with alternative figures obtained with the same method but with a 3 years reference period (2017-2019). The table also contains country rankings by relative excess mortality for both choices of reference period. Overall, from the comparison emerges how the shorter reference period results in slightly higher excess mortality estimates, with differences in country-specific estimates generally in the order of 1-2\%, with a maximum increase of 3.5\% fro South Korea, and a minimum increase of 0\% for the USA. Importantly, the change in reference period leaves relative country rankings largely unaffected, as is visible from the last two columns of the table. It is also worth noting that differences with the estimates obtained by Levitt et al. persist even when aligning the reference periods, showing that the use of more granular age classes is indeed an important factor. 

Zooming in on specific countries, it is interesting to note how our estimate for the USA remains constant at 15.6\% for both a 3-years reference period and 5-years one, keeping a distance of 0.9\% from the 14.7\% obtained by Levitt et al. The removal of 2015 and 2016 from the reference period did not change much in this case, which makes sense given that mortality levels in those two years were similar to 2017-2019, as visible from Fig. \ref{fig:non-EU} in the main paper. Moving our attention to Germany, the estimate for the country changes from +0.2\% with a 2015-2019 reference period, to +1.1\% using 2017-2019 as reference. This 0.9\% percent increase is in line with what is observed for most other European countries, for which estimates increase slightly without changing their overall character. This increase also brings the estimate for Germany closer to the 2.7\% obtained by Levitt et al. Nonetheless, a difference of 1.6\% between the two methods still persists despite the reference periods being the same. This is most likely due to the fact that, as is visible from the German age pyramid in Figure 1, the broad age classes used by Levitt et al. tend to have increasingly more weight in the older (and thus more mortal) age categories over time, thus leading to a slight underestimation in expected mortality for the Levitt method (as it implicitly assumes the composition of each age class to remain constant). Note that this does not only hold for Germany, but also for most other ageing European countries, and thus helps explain the moderate differences between the two sets of estimates for these countries. Nonetheless, the two methods remain largely in agreement with respect to both their overall assessment of excess mortality levels as well as the countries’ relative rankings.

\begin{table}[h!]
	\centering
	\begin{tabular}{lrrrrrr}
		\hline
		Country & 5Y Ref. & 3Y Ref. & 5Y\% & 3Y\% & 5Y Rank & 3Y Rank\\ 
		\hline
  	Bulgaria & 50,830 & 52,512 & +22.8\% & +23.7\% & 1 & 1\\ 
        USA & 920,731 & 923,133 & +15.6\% & +15.6\% & 2 & 2\\ 
        Czechia & 31,691 & 34,160 & +13.3\% & +14.5\% & 3 & 3\\ 
        Hungary & 26,653 & 28,650 & +9.8\% & +10.7\% & 4 & 5\\ 
        Lithuania & 8,000 & 10,470  & +9.6\% & +13.0\% & 5 & 4\\ 
		Croatia & 8,642 & 10,406 & +7.8\% & +9.5\% & 6 & 6\\ 
		Italy & 96,891 & 112,249 & +7.2\% & +8.4\% & 7 & 7\\ 
        UK & 71,808 & 85,957 & +5.6\% & +6.8\% & 8 & 8\\ 
        Portugal & 10,454 & 15,222 & +4.4\% & +6.5\% & 9 & 9\\ 
  	Netherlands & 14,175 & 17,924 & +4.4\% & +5.6\%  & 10 & 10\\ 
		Spain & 36,310 & 45,319 & +4.0\% & +5.1\% & 11 & 11\\ 
		Austria & 6,825 & 8,677  & +3.9\% & +5.0\% & 12 & 12\\ 
		Belgium & 6,470 & 10,571 & +2.8\% & +4.6\%  & 13 & 13\\ 
		France & 30,759 & 39,213  &  +2.4\% & +3.1\% & 14 & 15\\ 
        Switzerland & 2,256 & 4,785 & +1.6\% & +3.4\% & 15 & 14\\ 
		Canada & 8,910 & 11,515 & +1.5\% & +1.9\% & 16 & 16\\ 
		Germany & 4,098 & 23,775 & +0.2\% & +1.1\%  & 17 & 17\\ 
		Luxembourg & -113 & -87 & -1.2\% & -0.1\%  & 18 & 19\\ 
        Sweden & -2,681 & 29 & -1.4\% & +0.0\% & 19 & 18\\ 
  	Finland & -3,337 & -1,536  & -2.9\% & -1.3\% & 20 & 20\\ 
		Denmark & -3,552 & -2,539 & -3.1\% & -2.2\% & 21 & 22\\ 
  	Japan & -119,266 & -92,775  &  -4.1\% & -3.2\% & 22 & 24\\ 
		Ireland & -2,959 & -1,382 & -4.3\% & -2.1\% & 23 & 21\\ 
		Norway &  -4,418 & -3,060 & -5.1\% & -3.6\% & 24 & 25\\ 
		Hong Kong & -6,439 & -3,216 & -5.9\% & -3.1\% & 25 & 23\\ 
		Iceland & -289 & -190 & -5.8\% & -3.9\% & 26 & 26\\ 
        Taiwan & -26,232 & -19,255 & -6.8\% & -5.1\% & 27 & 27\\ 
        Australia & -25,628 & -20,041  &  -7.2\% & -5.7\% & 28 & 28\\ 
        New Zealand & -5,201 & -5,119 & -7.2\% & -7.0\% & 29 & 29\\ 
		South Korea & -62,034 & -44,076 & -9.1\% & -6.6\% & 30 & 30\\ 

		\hline
	\end{tabular}
	\caption{Excess mortality estimates obtained using 2015-2019 and 2017-2019 as reference periods, respectively. The table also contains country rankings (descending) by relative excess mortality for both choices of reference period.}
	
	\label{tab:reference}
\end{table}

\section{Recorded COVID-19 deaths}

Table \ref{tab:covid} contains year-specific official COVID-19 death counts for all country included in our analysis. Note that these numbers were not used in any of our analyses, which are solely based on all-cause mortality data. COVID-19 death counts are only provided here for comparison purposes. Also note that these figures should be interpreted with care, as testing and reporting strategies and capacities can vary heavily by country and over time (which is one of the reasons to generally prefer all-cause mortality to reported death tolls).
\begin{table}[ht!]
	\centering
	\begin{tabular}{lrrrrr}
		\hline
Country        & 2020   & 2021  & Total &  \\
		\hline
Australia      & 920    & 1523   & 2443   &  \\
Austria        & 7379   & 9318   & 16697  &  \\
Belgium        & 19638  & 8679   & 28317  &  \\
Bulgaria       & 7123   & 23405  & 30528  &  \\
Canada         & 15274  & 14684  & 29958  &  \\
Croatia        & 3860   & 8633   & 12493  &  \\
Czechia        & 11888  & 24444  & 36332  &  \\
Denmark        & 1256   & 2000   & 3256   &  \\
Finland        & 635    & 1292   & 1927   &  \\
France         & 63534  & 60631  & 124165 &  \\
Germany        & 47009  & 70683  & 117692 &  \\
Hong Kong      & 148    & 65     & 213    &  \\
Hungary        & 9537   & 29649  & 39186  &  \\
Iceland        & 28     & 9      & 37     &  \\
Ireland        & 2264   & 3820   & 6084   &  \\
Italy          & 73604  & 63643  & 137247 &  \\
Japan          & 3414   & 14979  & 18393  &  \\
Lithuania      & 1800   & 5598   & 7398   &  \\
Luxembourg     & 285    & 350    & 635    &  \\
Netherlands    & 11539  & 9421   & 20960  &  \\
New Zealand    & 25     & 26     & 51     &  \\
Norway         & 436    & 958    & 1394   &  \\
Portugal       & 6695   & 12193  & 18888  &  \\
South Korea    & 900    & 4725   & 5625   &  \\
Spain          & 53964  & 37311  & 91275  &  \\
Sweden         & 9706   & 5639   & 15345  &  \\
Switzerland    & 7528   & 4393   & 11921  &  \\
Taiwan         & 7      &  843   & 850    &  \\
United Kingdom & 93317  & 83496  & 176813 &  \\
United States  & 352004 & 467051 & 819055 &  \\
		\hline
\end{tabular}
	\caption{Officially recorded COVID-19 deaths by country and year.}
	\label{tab:covid}
\end{table}

\section{Excess mortality plots}
\label{sec:plots}

\begin{figure}[h!]
	\centering
 \includegraphics[width = 0.496 \textwidth]{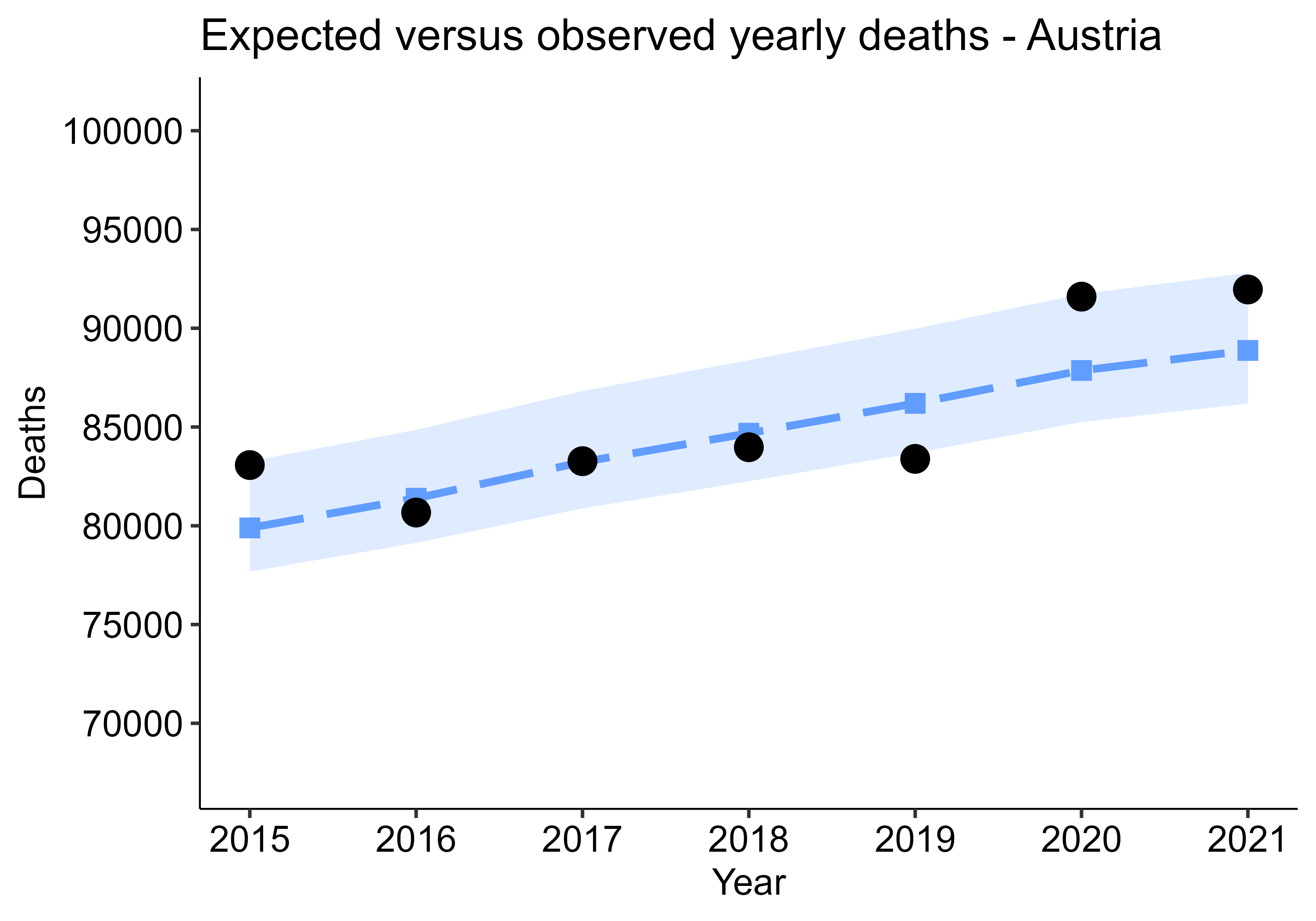} 
\includegraphics[width = 0.496 \textwidth]{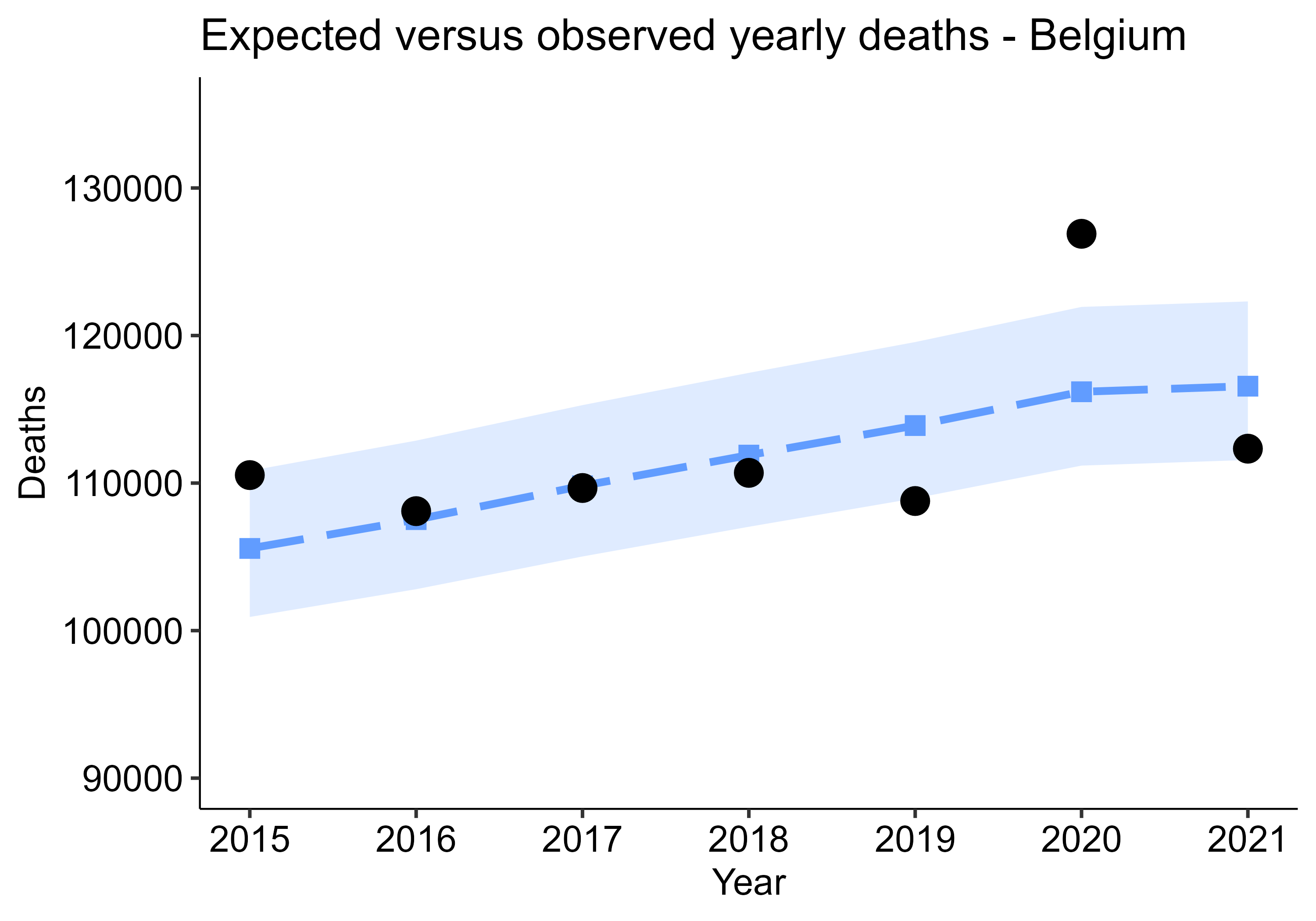} 
\includegraphics[width = 0.496 \textwidth]{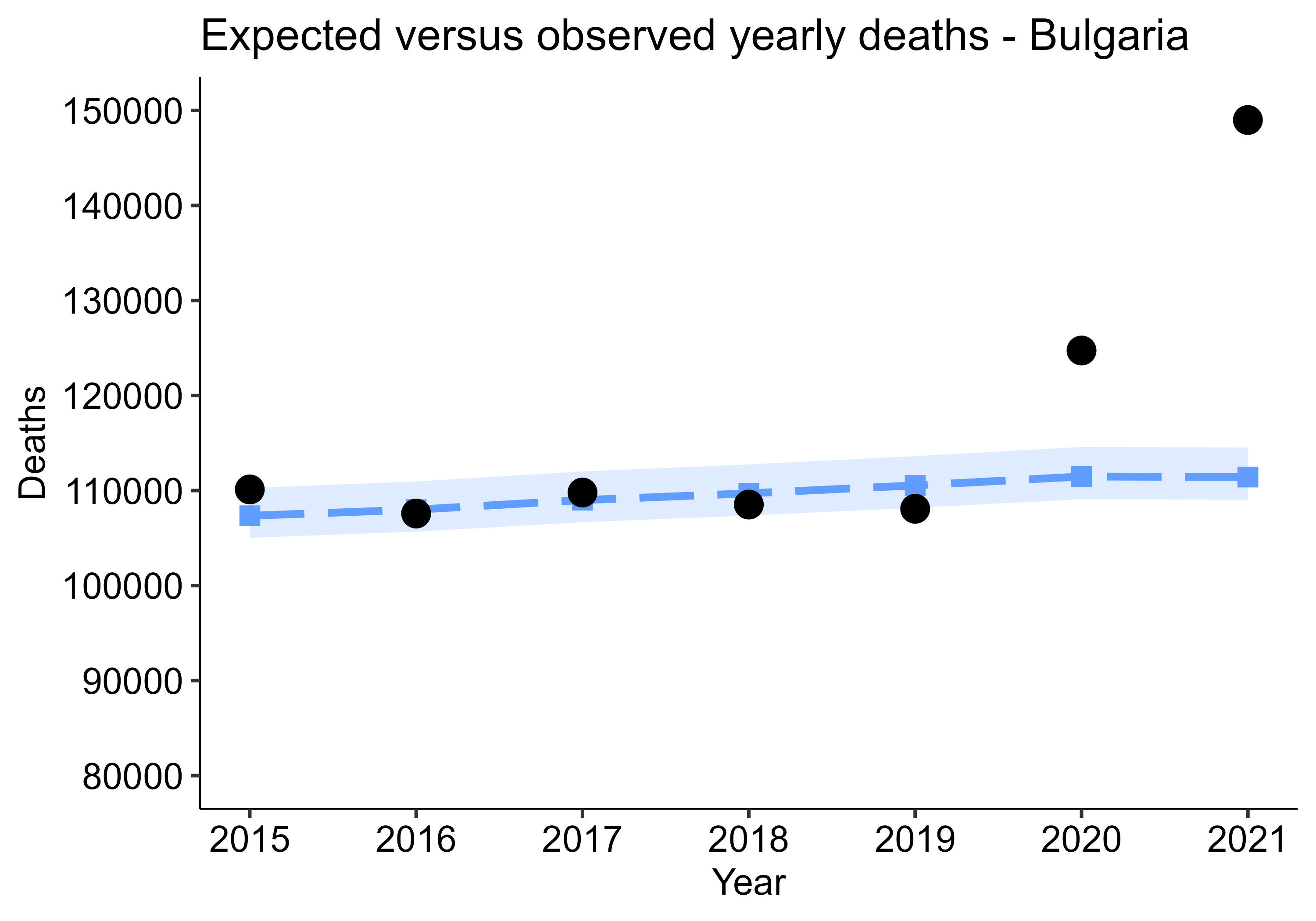}
\includegraphics[width = 0.496 \textwidth]{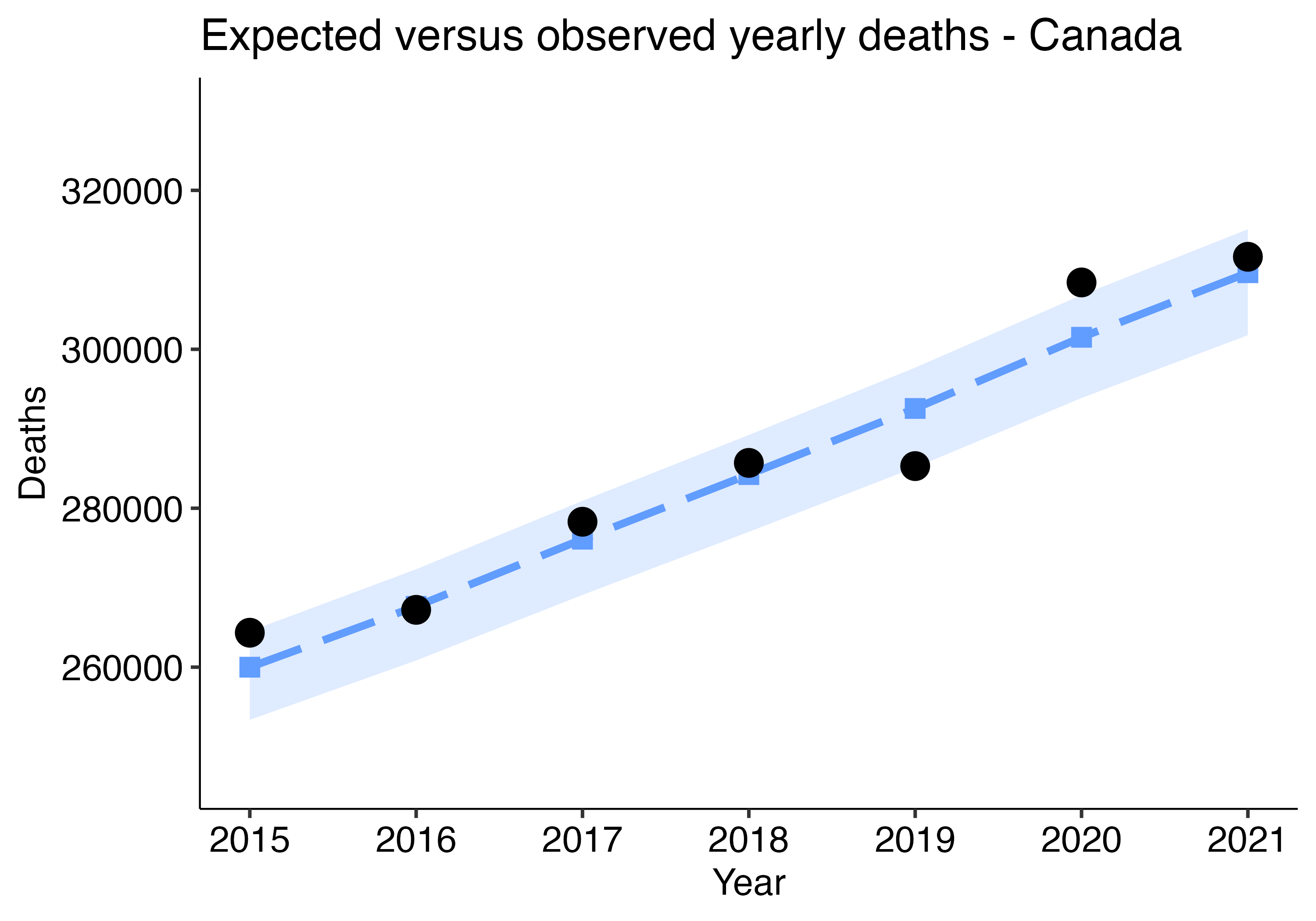}  
\includegraphics[width = 0.496 \textwidth]{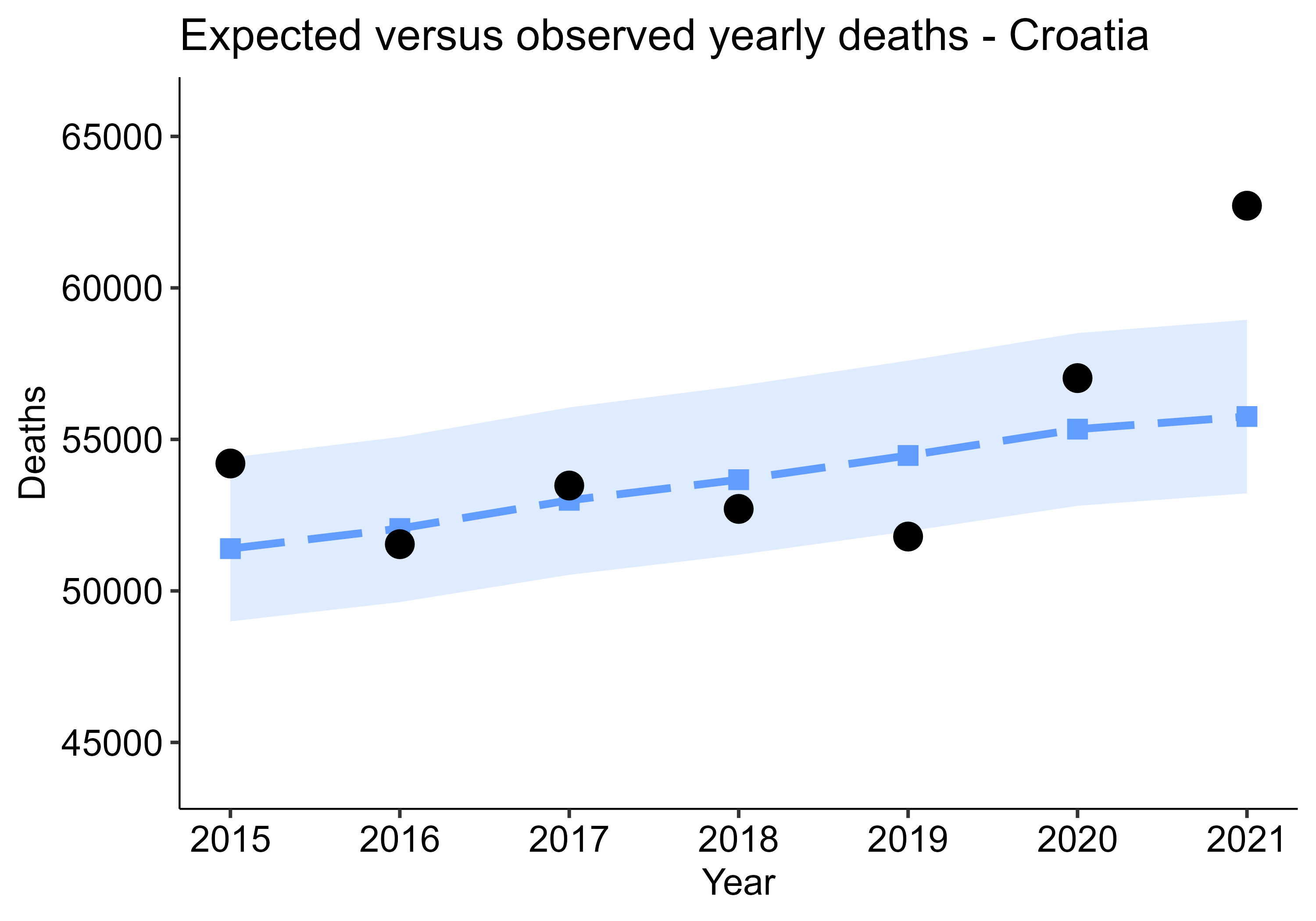} 
\includegraphics[width = 0.496 \textwidth]{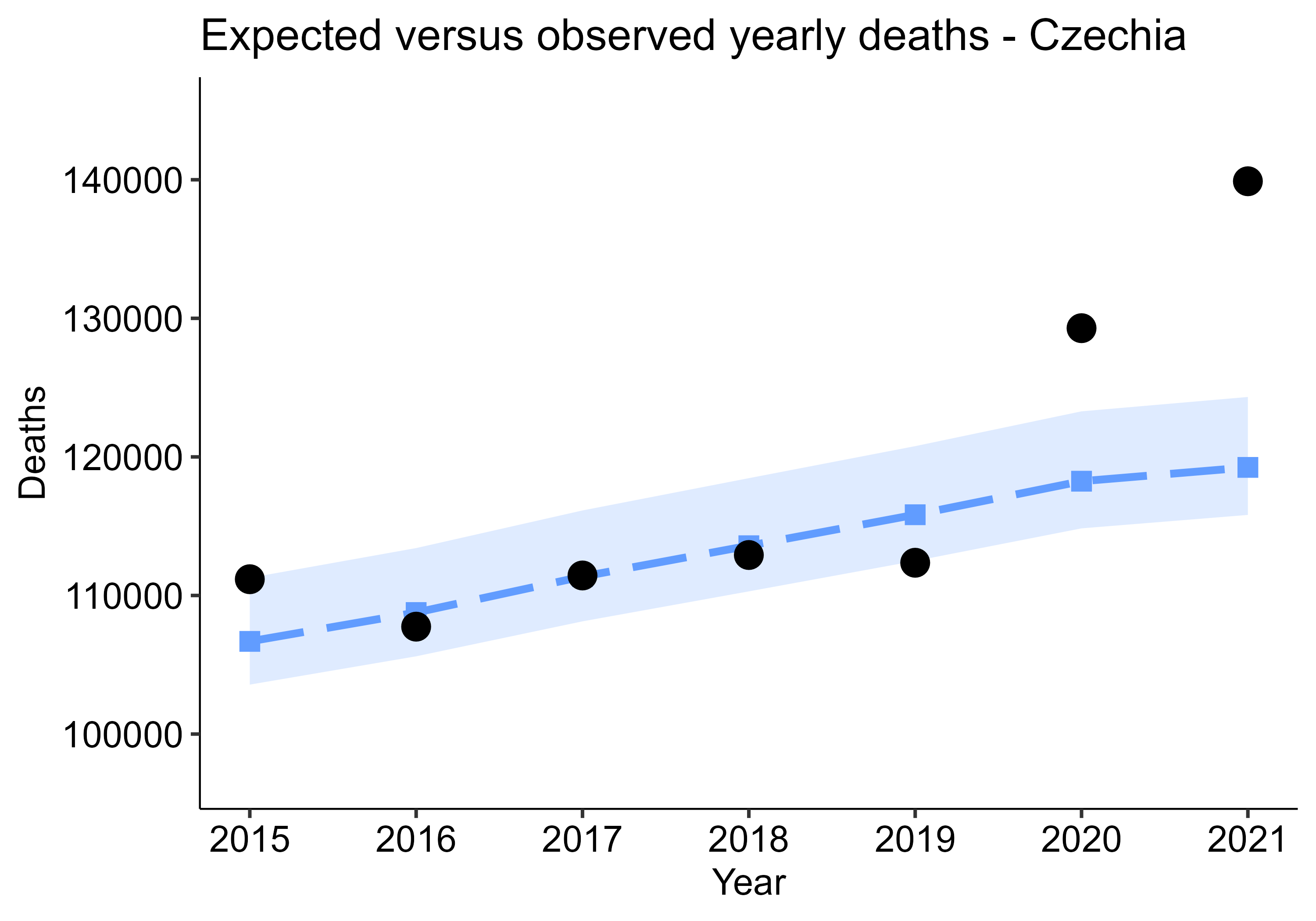}
\caption {Expected and observed mortality figures by calendar year for Austria, Belgium, Bulgaria, Canada, Croatia, and Czechia.}
\label{fig:sup1}
\end{figure}

\begin{figure}[]
	\centering

 \includegraphics[width = 0.496 \textwidth]{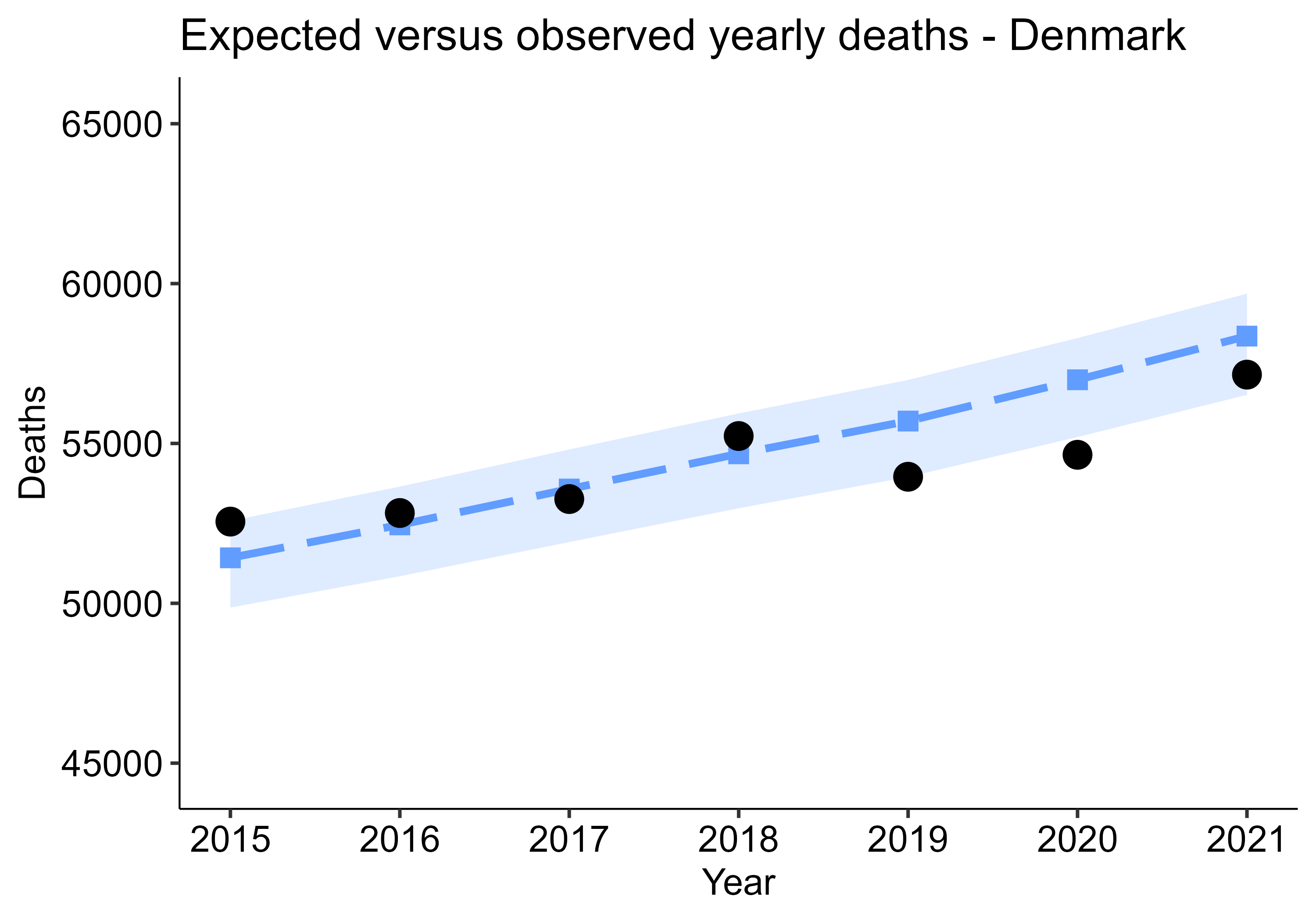} 
\includegraphics[width = 0.496 \textwidth]{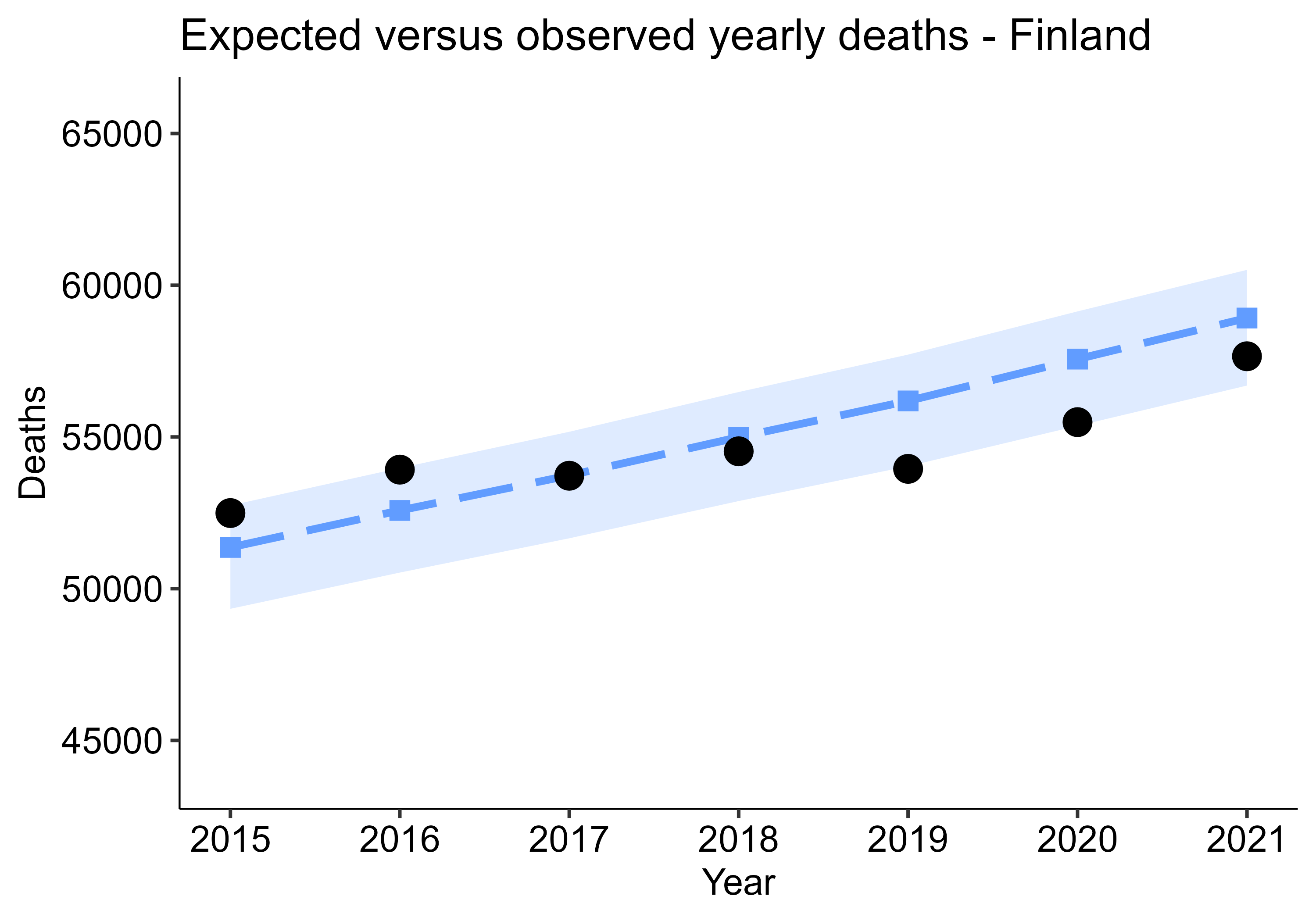}  
\includegraphics[width = 0.496 \textwidth]{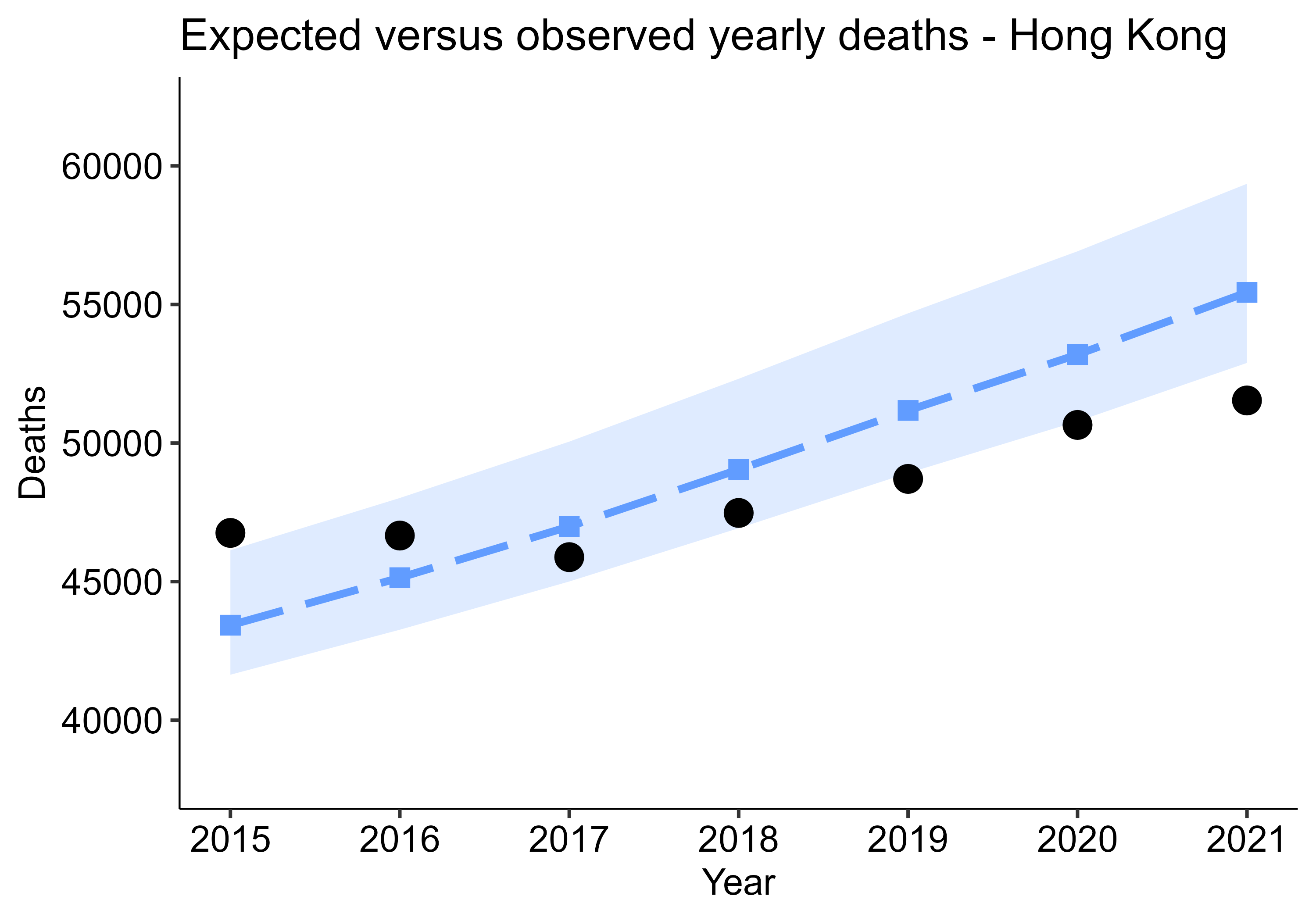}
\includegraphics[width = 0.496 \textwidth]{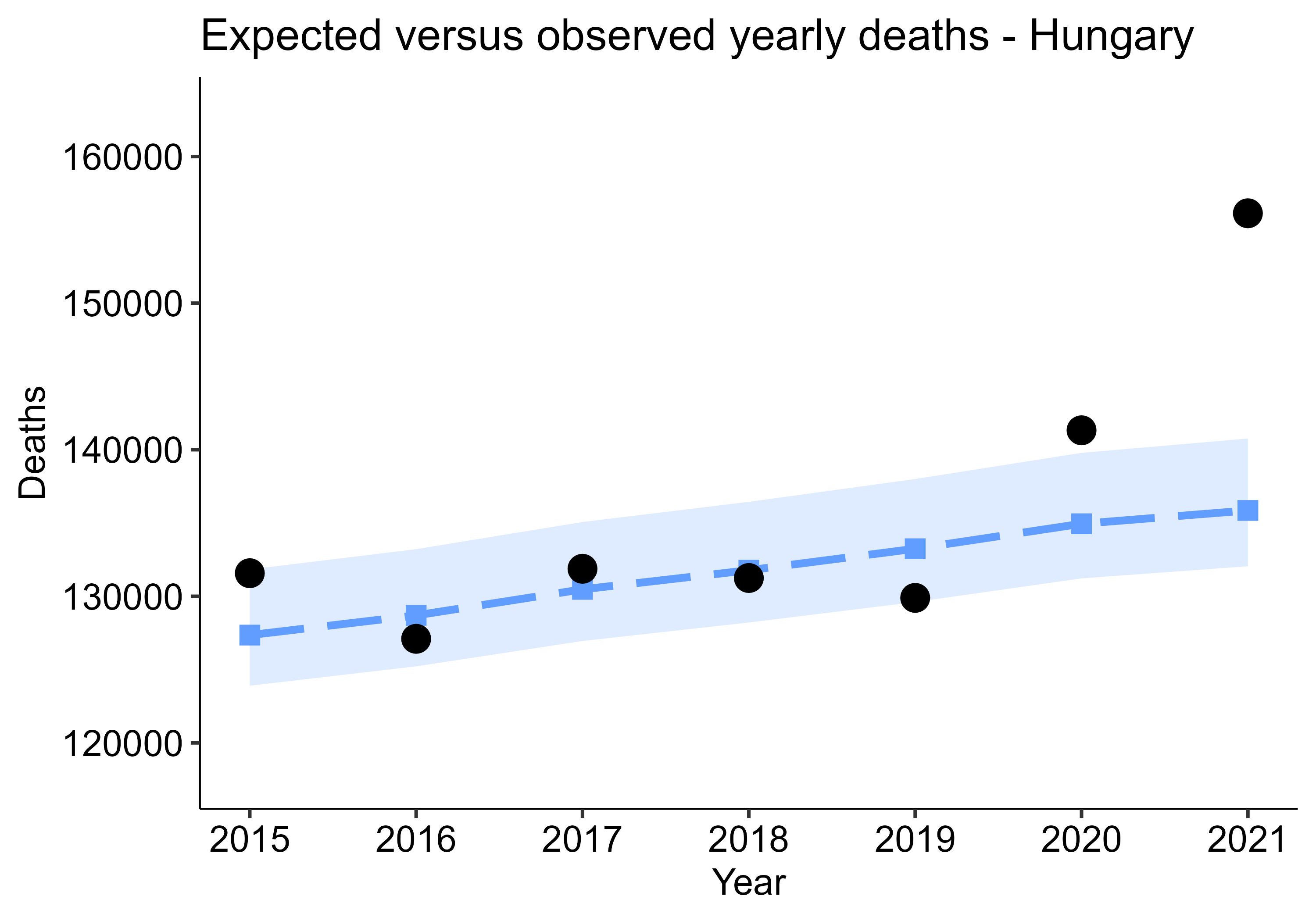} 
\includegraphics[width = 0.496 \textwidth]{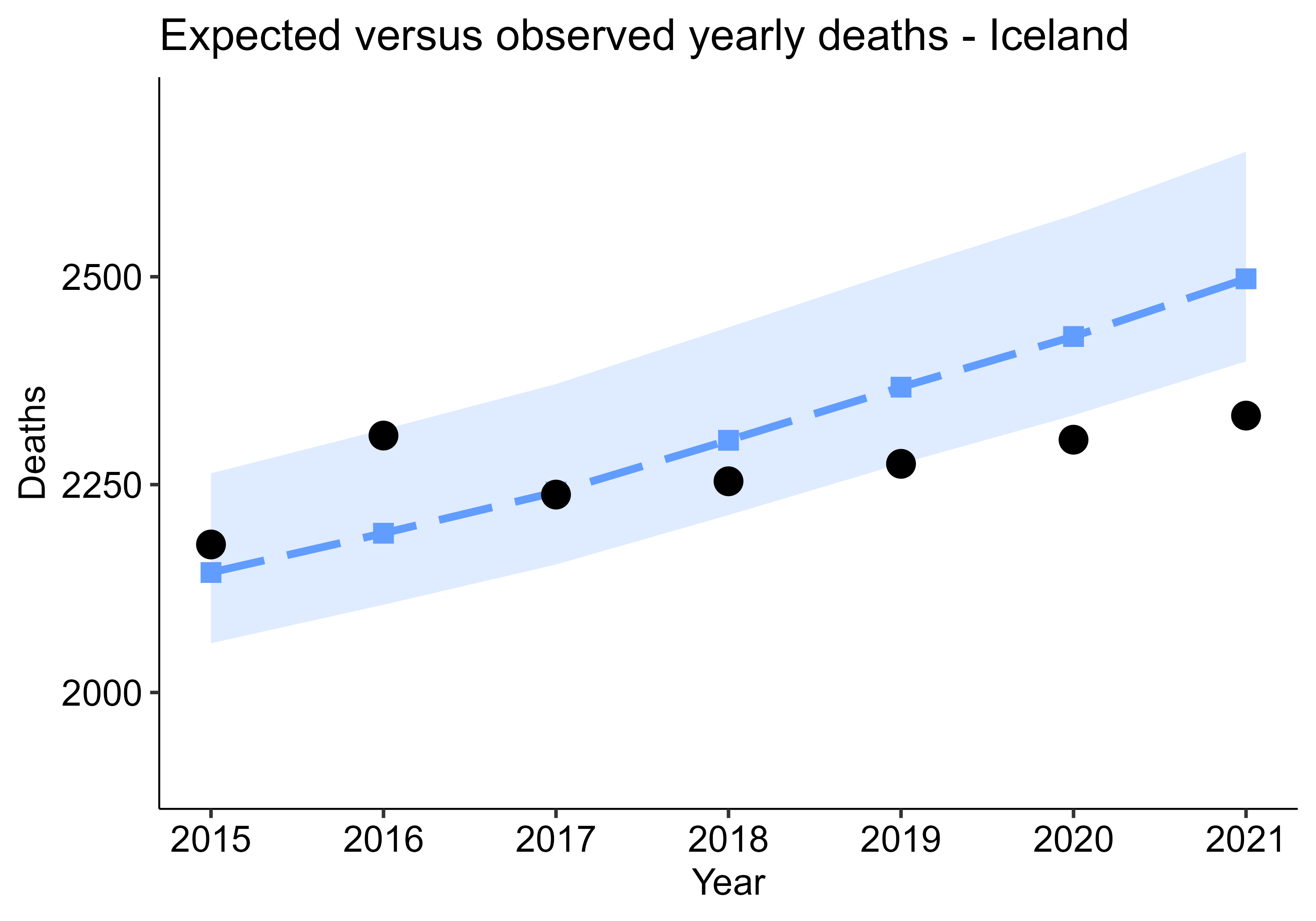} 
\includegraphics[width = 0.496 \textwidth]{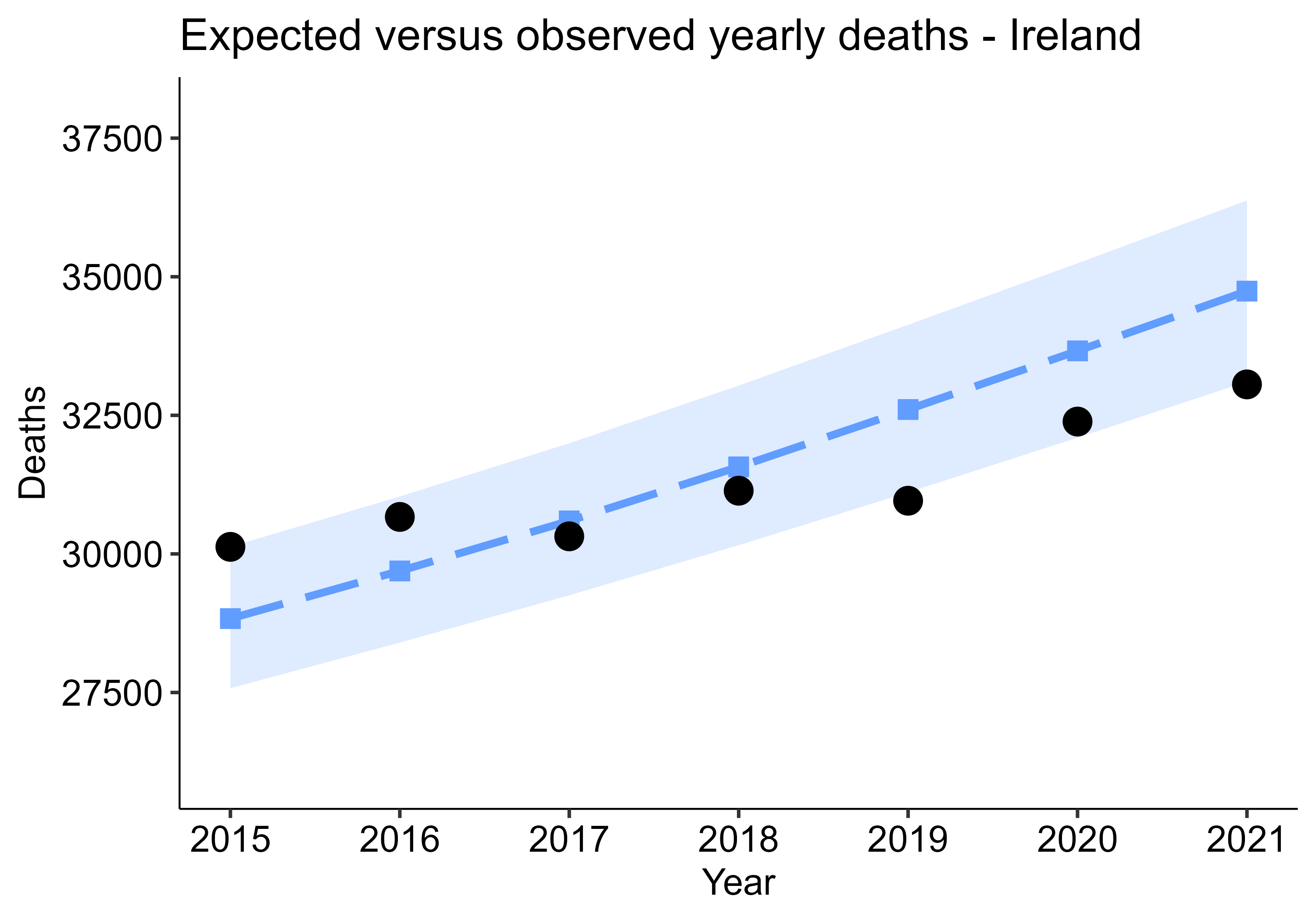} 
\caption {Expected and observed mortality figures by calendar year for Denmark, Finland, Hong Kong, Hungary, Iceland, and Ireland.}
\label{fig:sup2}
\end{figure}

\begin{figure}[]
	\centering 
\includegraphics[width = 0.496 \textwidth]{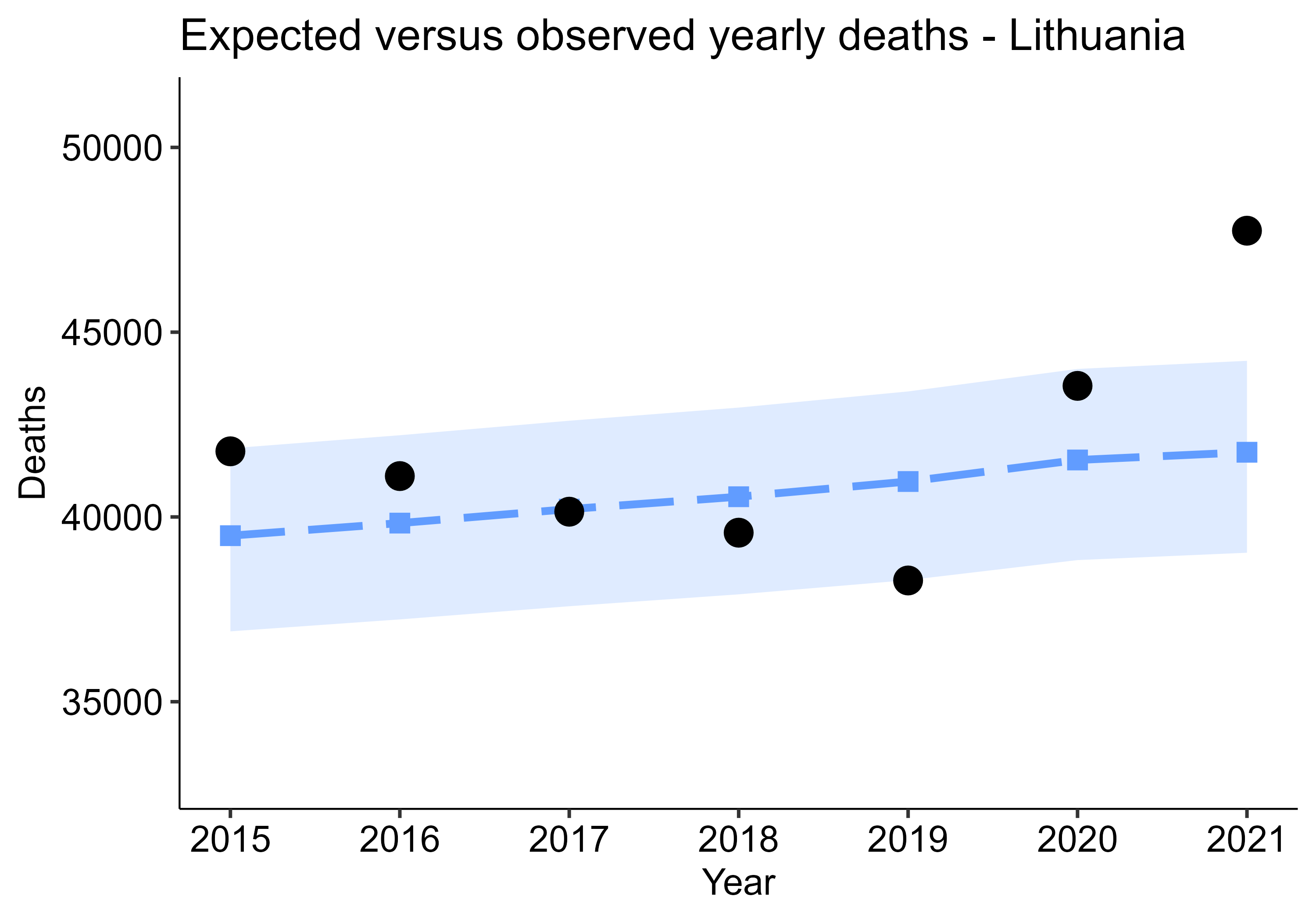} 
\includegraphics[width = 0.496 \textwidth]{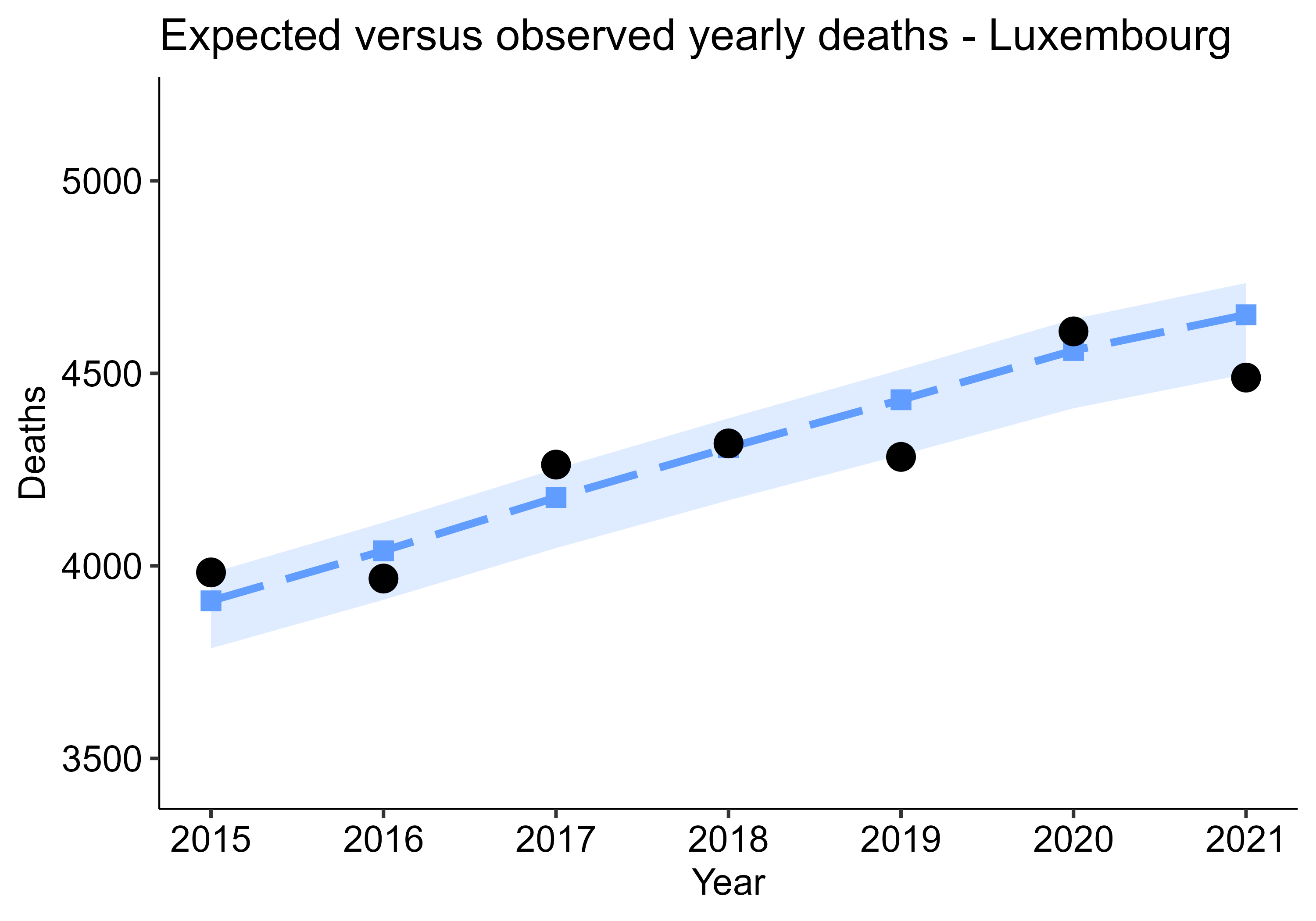}
\includegraphics[width = 0.496 \textwidth]{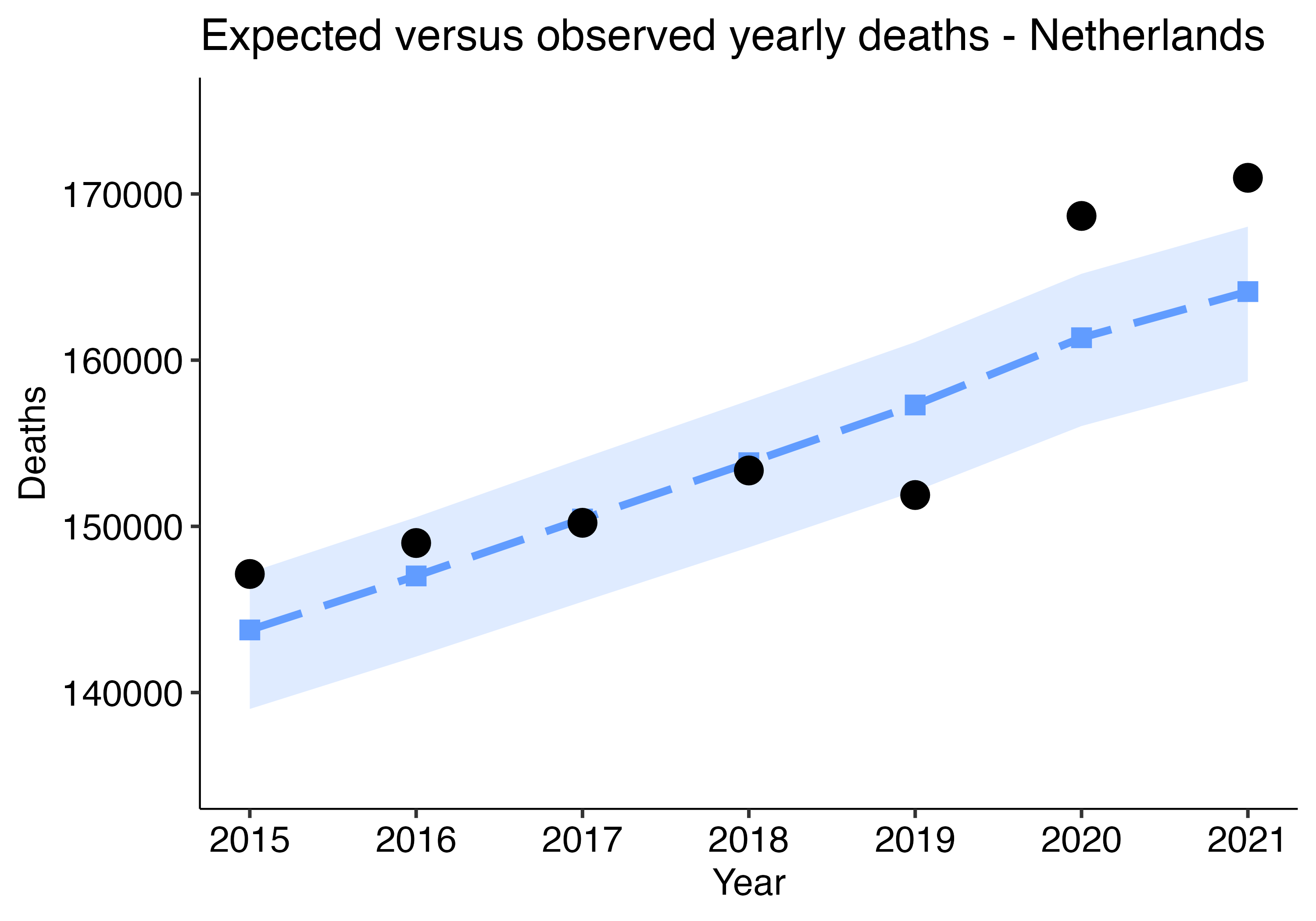} 
\includegraphics[width = 0.496 \textwidth]{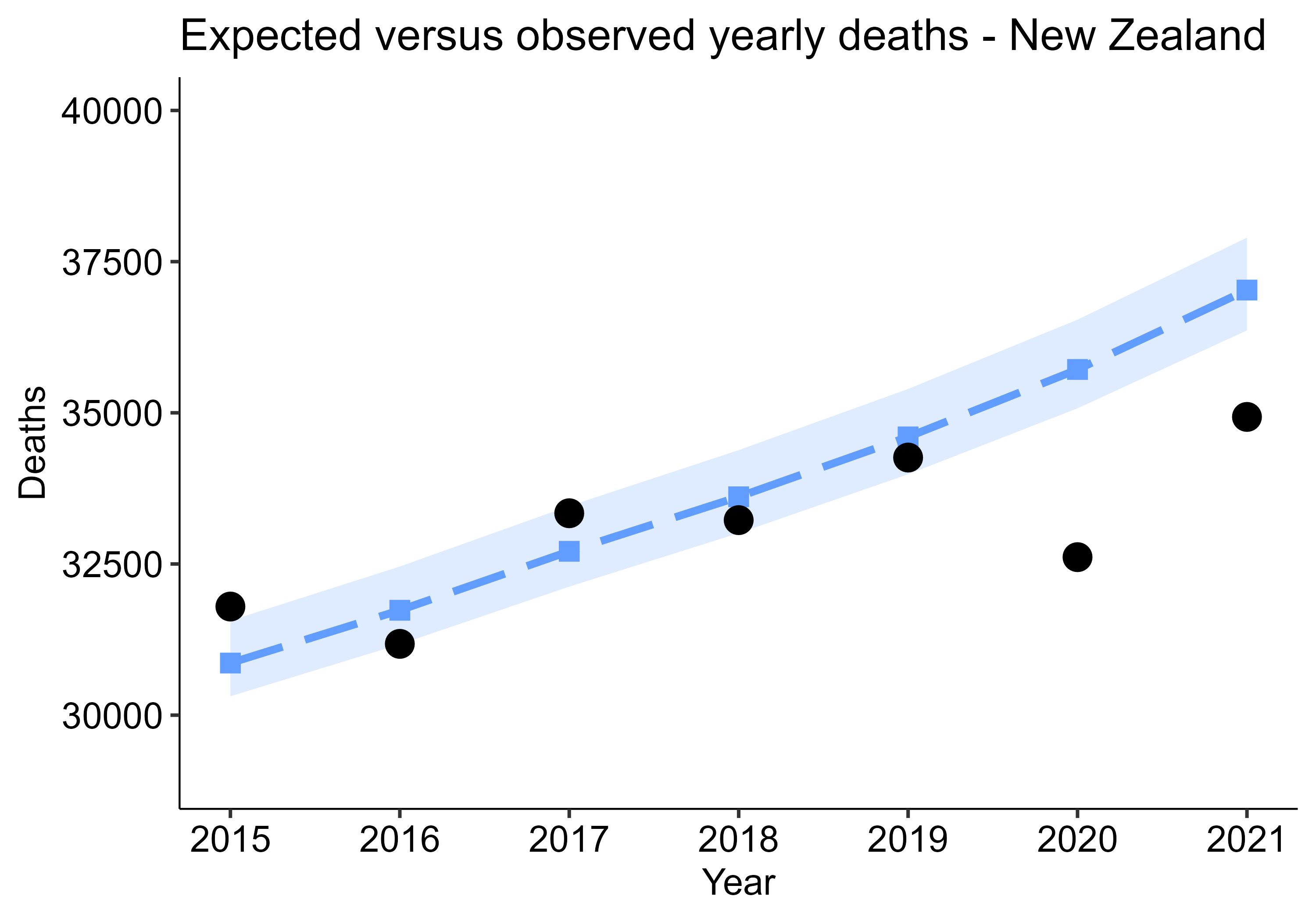} 
\includegraphics[width = 0.496 \textwidth]{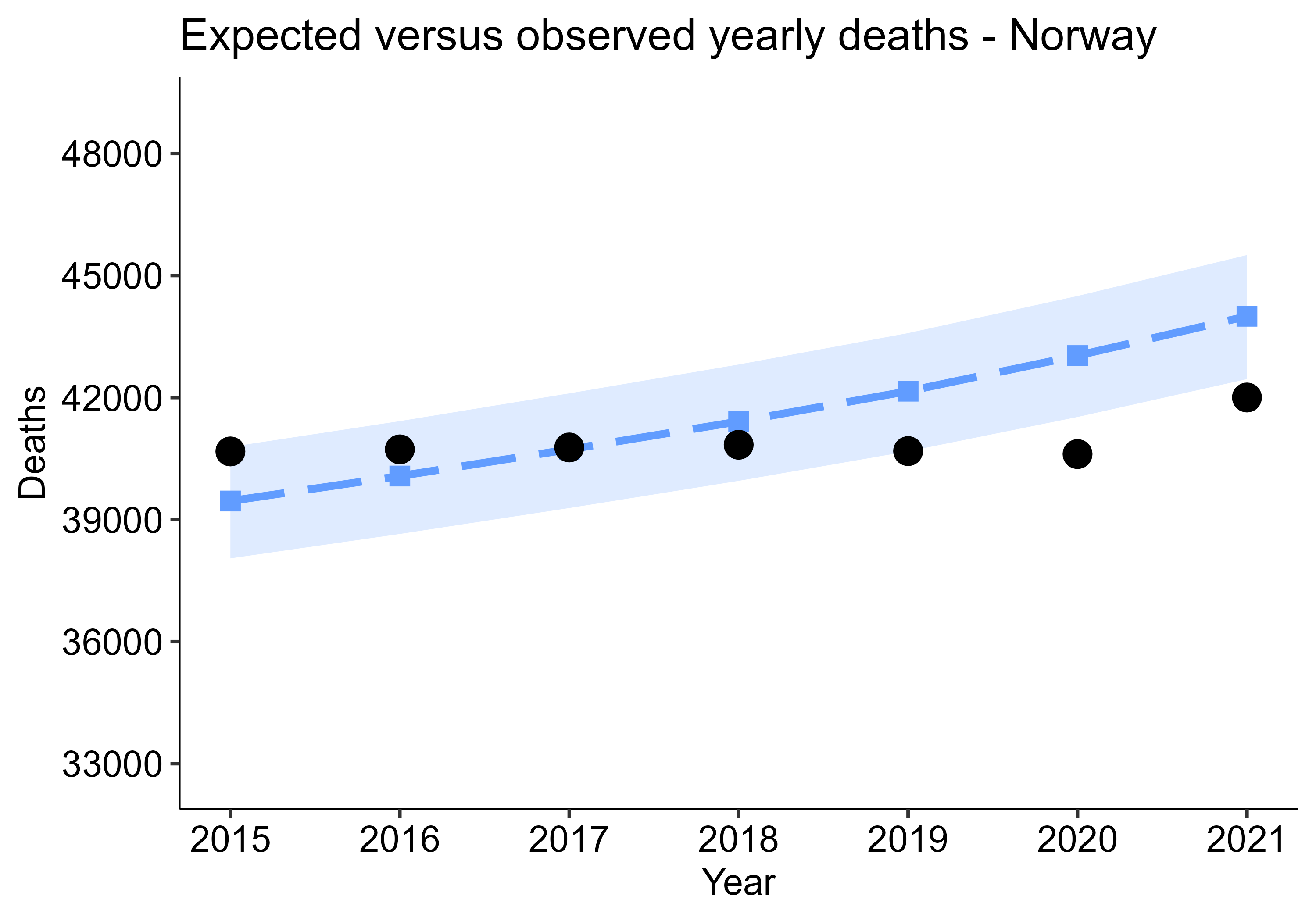} 
\includegraphics[width = 0.496 \textwidth]{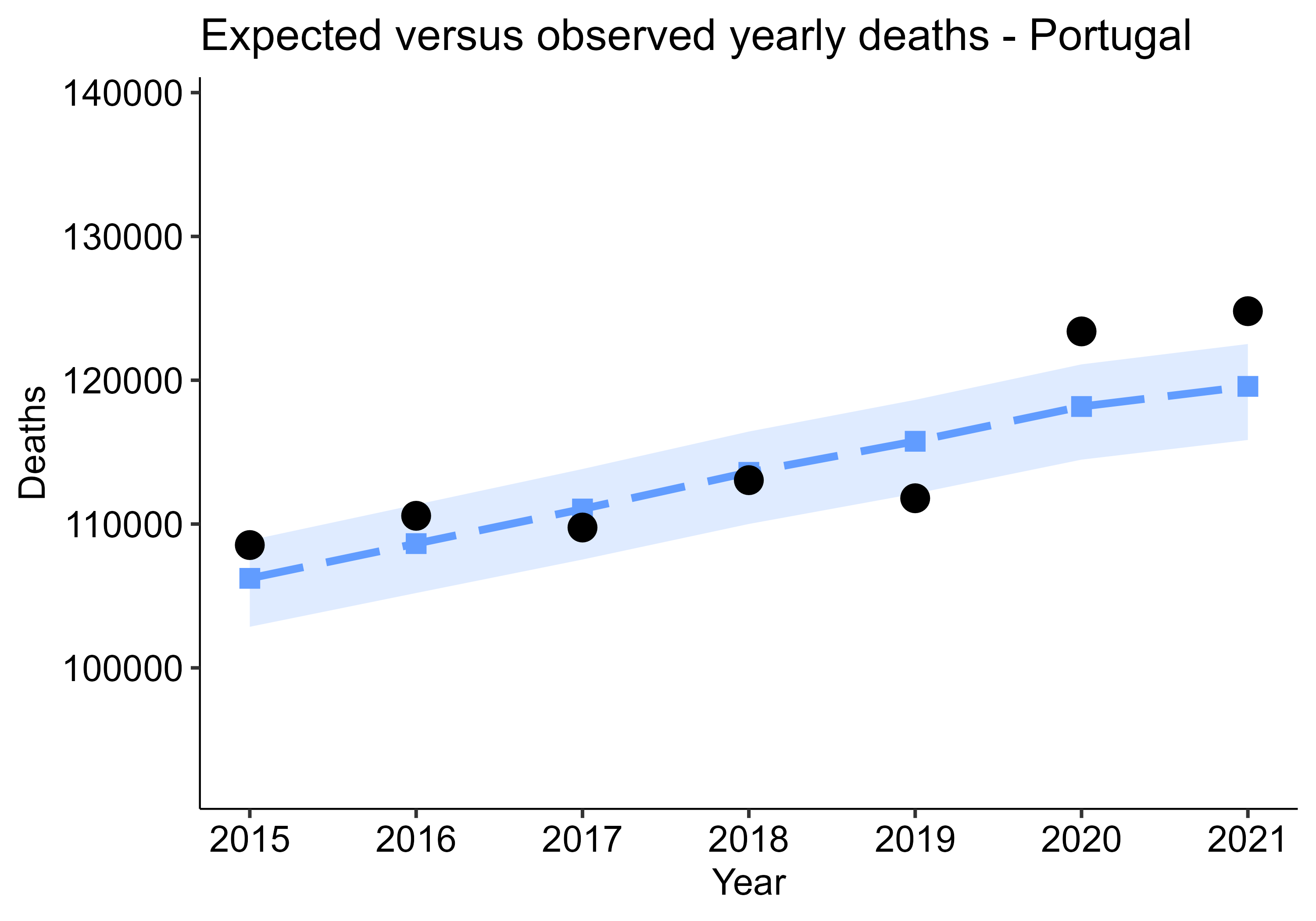} 
\caption {Expected and observed mortality figures by calendar year for Lithuania, Luxembourg, The Netherlands, New Zealand, Norway, and Portugal.}
\label{fig:sup3}
\end{figure}

\begin{figure}[]
	\centering
 \includegraphics[width = 0.496 \textwidth]{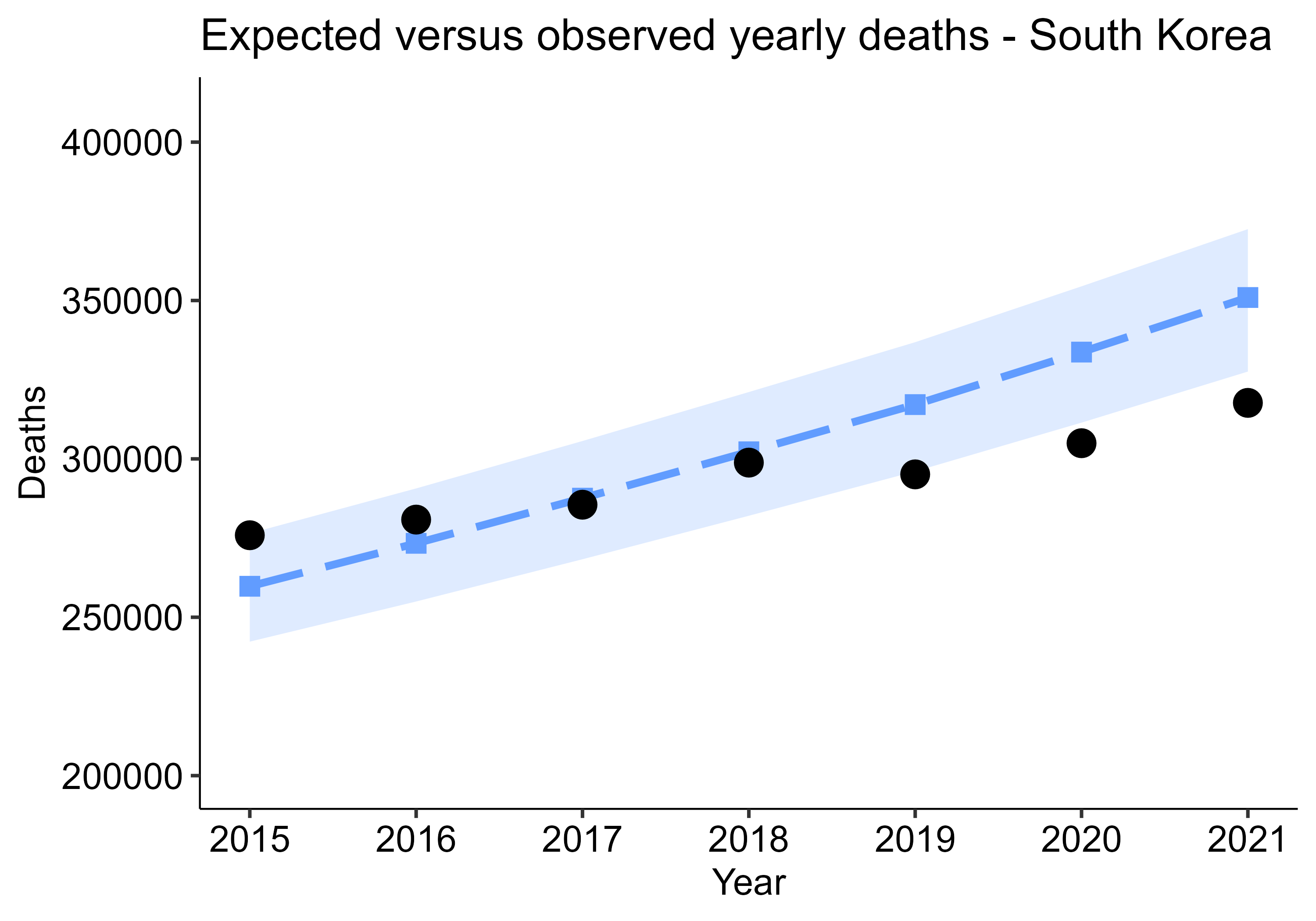} 
\includegraphics[width = 0.496 \textwidth]{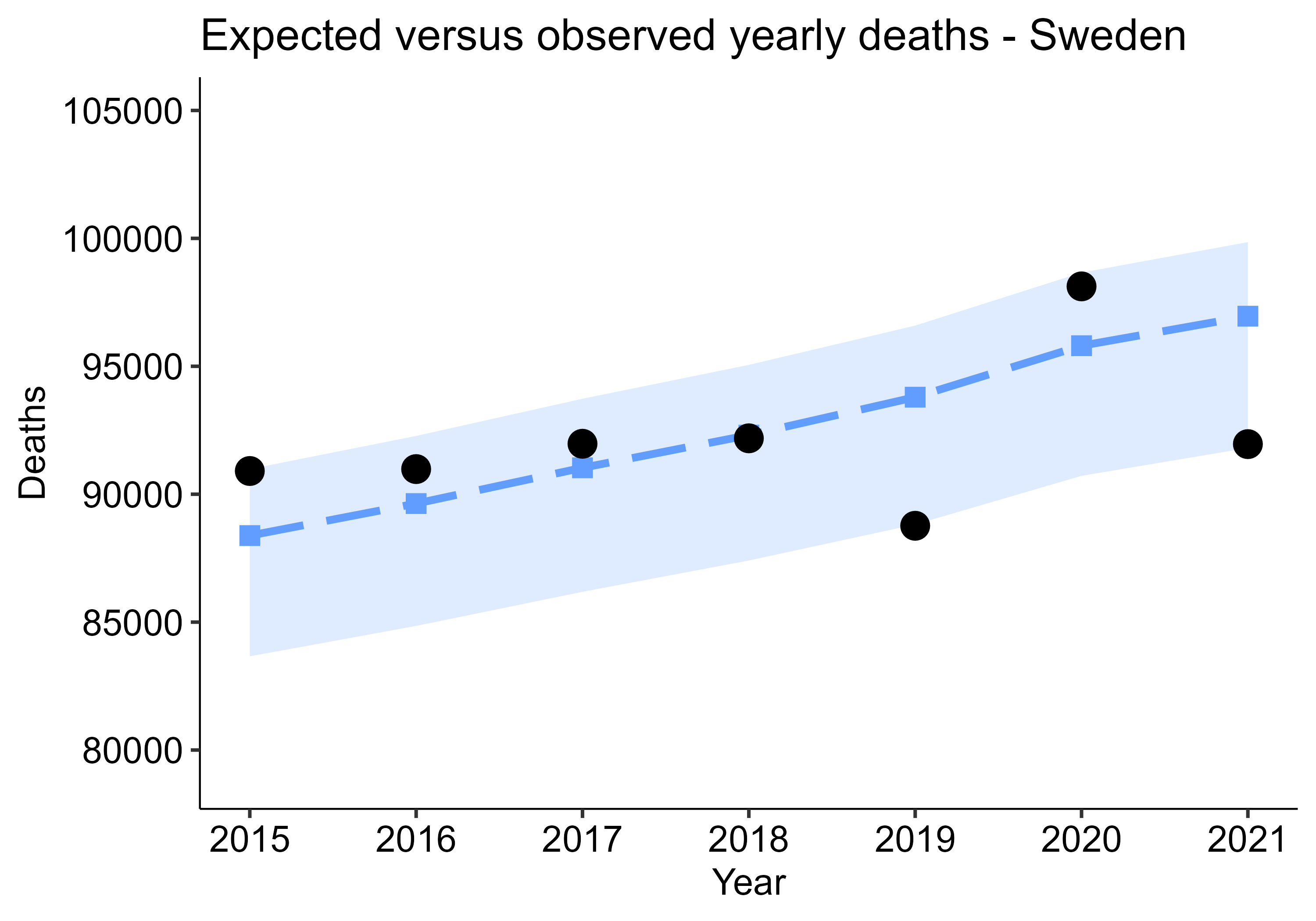}
\includegraphics[width = 0.496 \textwidth]{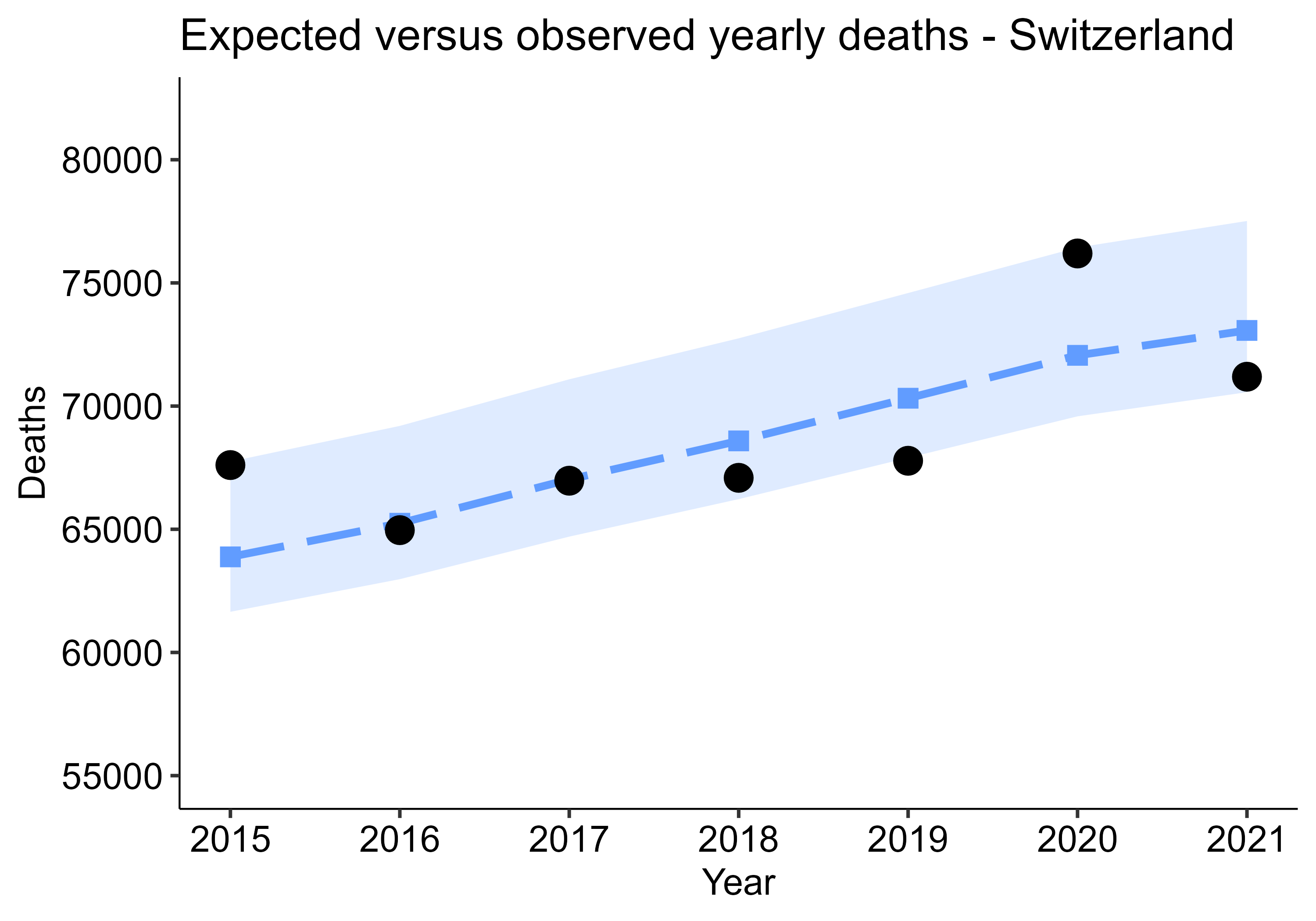} 
 \includegraphics[width = 0.496 \textwidth]{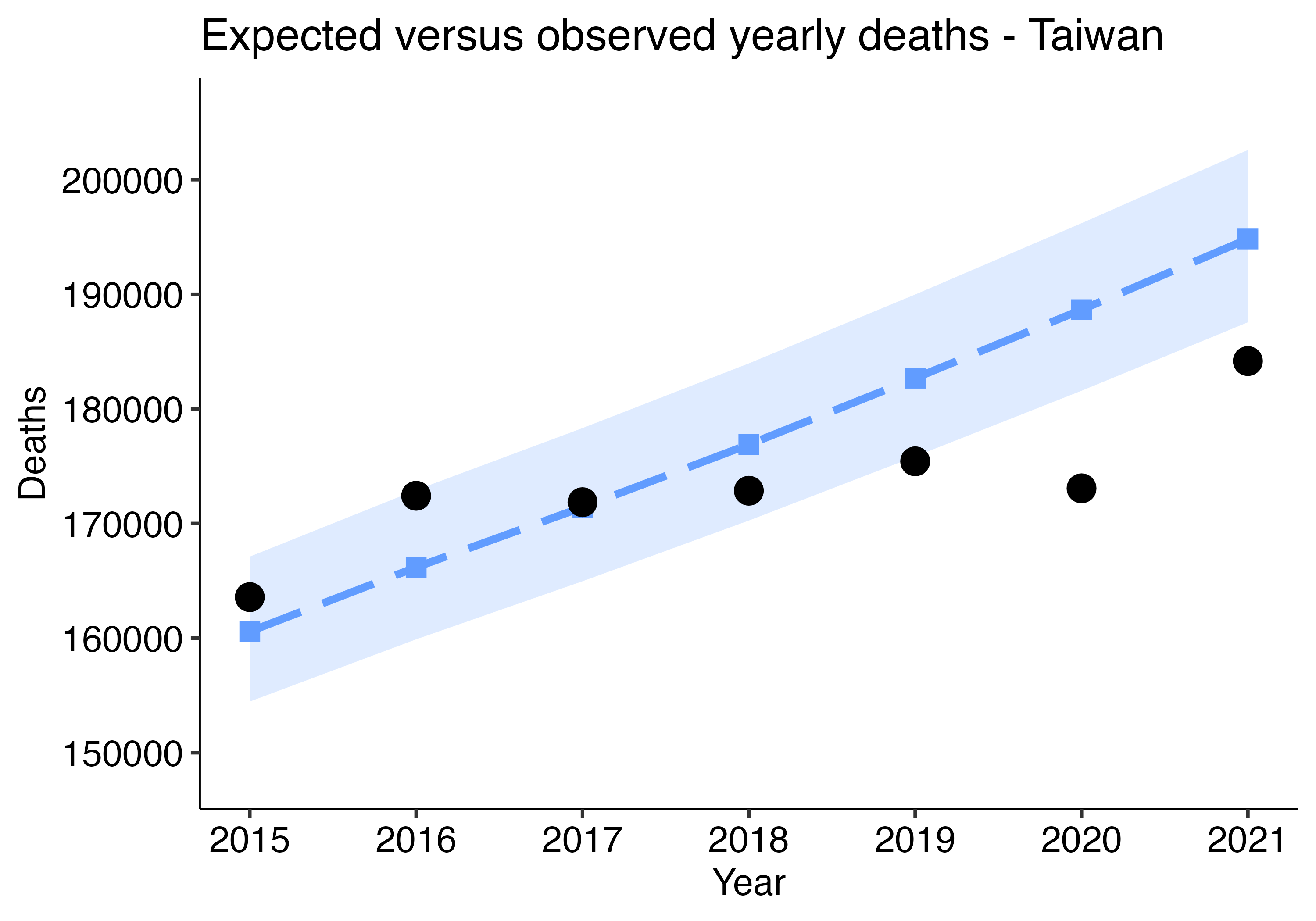} 
\caption {Expected and observed mortality figures by calendar year for South Korea, Sweden, Switzerland, and Taiwan.}
\label{fig:sup4}
\end{figure}

Figures \ref{fig:sup1}, \ref{fig:sup2}, \ref{fig:sup3}, and \ref{fig:sup4} display excess mortality plots equivalent to those displayed in Section \ref{sec:results} of the paper, for all countries included in the analysis for which the plot was not shown in the main text for reasons of brevity. In all plots, black dots indicate observed mortality in a given calendar year, while blue squares indicate estimated yearly expected mortality. Shaded bands represent the estimated expected mortality range. The plots are ordered alphabetically.

\vspace{1cm}

%\bibliographystylesupp{chicago}
%\bibliographysupp{literature}    

\end{document}